\documentclass[]{aa}  
\bibpunct{(}{)}{;}{a}{}{,}
\usepackage{txfonts}
\usepackage{graphicx}
\usepackage{natbib}
\usepackage{hyperref}
\usepackage[table]{xcolor}
\usepackage[flushleft]{threeparttable}
\begin{document}
    \nolinenumbers
   \title{Bypassing the static input size of neural networks in flare forecasting by using spatial pyramid pooling}

   \author{P. Vong
          \inst{1,}\inst{2}
          \and
          L. R. Dolla\inst{1}
          \and
          A. Koukras\inst{1,}\inst{2,}\inst{3}
          \and
          J. Gustin\inst{1}
          \and
          J. Amaya\inst{2}
          \and
          E. Dineva\inst{2}
          \and
          G. Lapenta\inst{2}\thanks{Deceased on 28 May 2024.}
          }

   \institute{Solar-Terrestrial Centre of Excellence - SIDC, Royal Observatory of Belgium, Avenue Circulaire 3, 1180, Brussels, Belgium\\
              \email{philippe.vong@oma.be}
                \and
                Center for Mathematical Plasma Astrophysics, Department of Mathematics, University of Leuven, KULeuven, Belgium
                \and
                Columbia Astrophysics Laboratory, Columbia University, MC 5247, 550 West 120th Street, New York, NY 10027, USA 
              }
   \date{Submitted September 15, 1996}
 
  \abstract
   {The spatial extension of active regions of the Sun (hence their associated images) can strongly vary from one case to the next. This inhomogeneity is a problem when using convolutional neural networks (CNNs) to study solar flares, as they generally use input images of a fixed size. Different processes can be performed to retrieve a database with homogeneous-sized images, such as coarse resizing, cropping, or padding of raw images. Unfortunately, key features can be lost or distorted beyond recognition during these processes. This can lead to a deterioration of the ability of CNNs to classify flares of different soft X-ray classes, especially those from active regions with structures of great complexity.}
   {This study aims to implement and test a CNN architecture that retains features of characteristic scales as fine as the original resolution of the input images.}
   {We compared the performance of two CNN architectures for solar flare prediction. The first one is a traditional CNN with convolution layers, batch normalization layers, max pooling layers, and resized input, whereas the other implements a spatial pyramid pooling (SPP) layer instead of a max pooling layer before the flattening layer and without any input resizing. Both were trained on the Spaceweather HMI Active Region Patch (SHARP) line of sight magnetogram database, which was generated from data collected by the Helioseismic and Magnetic Imager on board the Solar Dynamics Observatory from May 2010 to August 2021 only using images within 45\degr of the central meridian of the Sun. We also studied two cases of binary classification. In the first case, our model had to distinguish active regions producing flares in less than 24h of class $\ge$C1.0 from active regions producing flares in more than 24h or never. In the second case, it had to distinguish active regions producing flares in less than 24h of class $\ge$M1.0 from active regions producing flares in more than 24h or never, or flares in less than 24h but of class <M1.0. The impact of the use of a score-oriented loss (SOL) function optimizing the true skill statistics (TSS) metric instead of a binary cross-entropy (BCE) loss function is also studied and discussed in this work.}
   {Our models implementing an SPP layer and trained using a BCE loss function outperform the traditional CNN models, with an average increase of 0.1 in TSS and 0.17 in precision metrics when predicting flares $\ge$C1.0 within 24h. However, their performances degrade sharply along the other models studied in this paper when trained to classify images of $\ge$M1.0 flares.}
   {We prove the higher efficiency of a CNN model that includes an SPP layer in predicting solar flares. The degradation of prediction performance of this model when the images of active regions producing a C class flare are classified as negative may be attributed to its success in identifying features that appear in active regions only a few hours before the flare, independent of their soft X-ray class. The development of explainable artificial intelligence tools adapted to this architecture in future projects will be interesting for the study of solar flare-triggering mechanisms.}


   \maketitle 
%

\section{Introduction}
  The Sun is a star that presents a complex magnetic field that can become intertwined and reconnect under many circumstances. Whenever this happens, energy is released through electromagnetic radiation and particles. Such an event is called a flare and typically occurs in the so-called active regions \citep[AR;][]{Toriumi2019}. A proxy of the amount of energy released during flares is routinely provided by measuring the soft X-ray flux with the Geostationary Operational Environmental Satellite (GOES). From the peak flux, flare classes are defined on a logarithmic scale: A, B, C, M, and X. Intermediate steps within the class can be used by appending a value to the letter, still on a logarithmic scale. Whenever electromagnetic radiation and particles reach the near-Earth space environment, they can negatively impact living beings, electronic devices, infrastructures, and cause many more disruptions. Several reports have described impacts on power grids, radio communication, GPS, and climate change linked with intense solar activity \citep{Boteler2003,Yasyukevich2018,Gulati2019,Airapetian2020}. Therefore, to prevent further damage and react adequately, it is necessary to be able to predict when solar flares will happen. \\
   
   Solar flare forecasters usually try to predict events through visual recognition with data, such as magnetograms and extreme ultraviolet images. Unfortunately, it is very time-consuming and often inaccurate due to the lack of knowledge on the triggering mechanism behind solar flares. Many researchers and engineers have tried to develop prediction models to assist the forecasting work based on the technology available at the time. Early examples include statistic-based models from \citet{Song2009}, \citet{Mason_2010}, and \citet{Bloomfield_2012}. Based on statistics extracted and calculated from the magnetic field around sunspots of an active region, these methods had relative success but were limited in their prediction capability. With the emergence of machine learning, solar flare prediction models have quickly converged toward machine learning-based methods. The different prediction methods that have been explored include artificial neural networks \citep{FernandezBorda2002,2008Wang,2013Ahmed}, support vector machines \citep{Qahwaji2007,Bobra_2015}, K-nearest neighbors \citep{2007Li,2015Winter}, and ensemble methods \citep{2017Nishizuka,2021Ribeiro}. While machine learning methods show great success, they hold the same flaw as statistical methods. They can recognize patterns among the data and make predictions, but they need a human operator to feed them specific features or input that has been manually extracted from the raw data.\\
   
   During the past few years, a new branch of machine learning called deep learning (DL) has emerged and fixed this flaw by presenting several ground-breaking methods. Deep learning algorithms may be regarded as a mathematically complex evolution of machine learning, while its models are mainly structured as a multilayered neural network. A specific characteristic of deep learning methods is their ability to learn to autonomously recognize features from raw data and make predictions based on them. Convolutional neural networks \citep[CNNs;][]{LeCun2015} and long short-term memory \citep[LSTM;][]{hochreiter1997long} are two notable and prominent methods in this branch. Convolutional neural networks are especially famous in the DL field due to their ease to be used and their impressive results \citep{alexnet}. The main characteristics of CNNs are their ability to extract features with convolution layers and reduce the dimension and size of the input with pooling layers (several terms specific to CNNs and DL are defined in the beginning of section 3). Many innovations have emerged from CNNs, such as YOLO \citep{redmon2016look}, and are now mainly used in computer vision and image recognition.\\
   During the past years, there has been an increase in the use of CNNs\citep{ABED20212544,Huang_2018,Li2020,2021Yi} and LSTM \citep{2019Liu,2020Wang,Guastavino2022} to forecast solar flares \citep{flarecast}. Although studies have presented relatively good results, there have been many incoherent variations in prediction skill scores. This can be explained by the difference in realization and a lack of standardization among studies. \\
   
   While studying the methodology of various machine learning-based studies, we observed the following inconsistency. If we put aside the difference between the architectures and DL methods used, the first discernible problem is a divergence in the data-labeling process. Deep learning models usually try to predict flares with a GOES class $\ge$C1.0 or $\ge$M1.0 within 24 hours before the eruption time through a binary classification. However, to establish if an image is within 24 hours of a flare, an exact flaring time needs to be declared, whether it is the start time, the peak time, or the end time. Furthermore, while images need to be labeled based on the intensity of future flares in order to analyze the prediction performance of their models, there has been a discrepancy between studies. Some studies use the strongest flare within 24 hours \citep{Guastavino2022}, while other studies use some other labeling methods \citep{Li2020}. Another cause of the difference in results between DL models is the image selection conditions. Some papers claim the use of images within 45\degr of the central meridian \citep{Li2020, 2020Bhattacharjee}, whereas others choose 30\degr \citep{Huang_2018, Sun_2023} or are without any condition regarding the image position \citep{Guastavino2022}. It is especially impactful when there is a notable projection effect on the images, as the active region comes closer to the solar limb. Apart from the difference between sources and image types, the time range from which datasets are extracted can also impact the skill scores of the models. The Sun presents a cycle of 11 years, and depending on the phase of the cycle, there can be a notable difference in the shape of sunspots, the intensity of flares, and their positions. \\
   Building a dataset using the correct criteria and labeling is essential, especially while ensuring that no duplicates are found across multiple sub-datasets, in other words ensuring that its parsimony is respected. It is common knowledge to correctly split the dataset between a training, validation, and test sub-dataset in order to train a DL model. If the same image is in multiple sub-datasets, it may invalidate the evaluation of the model or even its training. Moreover, this problem is particularly sensitive in solar flare prediction because active regions are often observed multiple times during their lifetimes and can appear dozens to hundreds of times in datasets while presenting no apparent change. To respect the parsimony of datasets, one cannot simply shuffle and split the dataset. Confirming the separation of images of the same active region in multiple splits is imperative for correct training. Unfortunately, many studies do not ensure this \citep{Park2018, 2021Yi}, and some do not even distinguish between validation and test datasets \cite{Li2020}.\\
   Apart from the differences between datasets, there are also contrasts in evaluation methods. Analyzing the result of a model after a single training does not give any information about its stability and reproducibility. Therefore, to evaluate an architecture with a specific set of parameters, k-fold cross-validation is used to obtain reliable statistics by training a model k times on different splits of the dataset.\\
   The use of DL methods is still relatively new in solar flare forecasting. Unfortunately, there is no general agreement between forecasters or mainstream tools to create standardized datasets. Some custom-made datasets for DL model training are emerging \citep{Angryk2020,Boucheron2023} and look promising, but they are still recent and not fully studied. \\
   
   Due to the fixed number of neurons in fully connected layers, traditional CNNs are fed data with a fixed size regardless of the use of mini-batches. The most common way to meet this criterion is by resizing the input to the desired size. The resized format is usually a squared image with an edge size between 100 and 200 pixels. This format allows for faster training and prediction, but a lot of information can be lost during the process. This is especially prevalent with strong solar flare observations. These images tend to be larger and present complex small structures.\\
   In this paper, we attempt to present a solution to this problem by implementing a spatial pyramid pooling \citep[SPP;][]{He2014_spp} layer before the flattening layer in a binary classification model based on \citet{Li2020}. We call this architecture the SPP-CNN. We evaluate this architecture by training a model with a traditional max pooling layer instead of the SPP layer. It will serve as a control as we limit the difference between both architectures. We call this architecture the Trad-CNN. Following this, we analyze our results and discuss the future and the flaws of our study.
\section{Data}    

    \begin{table*}
        \caption{Dataset distribution during each filtering step.}             
        \label{table:1}      
        \centering          
        \begin{tabular}{c c c c c c c c c c}
        \hline\hline     
        Filtering & Category & Total & X & M & C & NF & FX & FM & FC\\
        \hline                    
           No filtering & Images count & 315 374 & 932 &  7 892 &  45 425 &  196 408 &  1 049 &  8 666 &  55 002\\
            & Images percentage & … & 0.3\% & 2.5\% & 14.4\% & 62.28\% & 0.33\% & 2.75\% & 17.44\%\\
            & AR count & 1421 & 27 & 171 & 651 & 1311 & 15 & 112 & 536\\
            & AR total percentage & … & 1.9\% & 12.03\% & 45.81\% & 92.26\% & 1.06\% & 7.88\% & 37.72\%\\ \\
           
           Post size filtering & Images count & 290 006 & 892 & 7 736 & 44 094 & 174 575 & 1004 & 8507 & 53 198\\  
            & Images percentage & … & 0.31\% & 2.67\% & 15.2\% & 60.2\% & 0.35\% & 2.93\% & 18.34\%\\
            & AR count & 1276 & 27 & 166 & 636 & 1 143 & 15 & 112 & 525\\
            & AR total percentage & … & 2.12\% & 13.01\% & 49.84\% & 89.58\% & 1.18\% & 8.78\% & 41.14\%\\ \\
           
           Post position filtering & Images count & 141 378 & 287 & 2 686 & 17 094 & 91 157 & 324 & 3 333 & 26 497\\  
            & Images percentage & … & 0.2\% &1.9\% &12.09\% &64.48\% &0.23\% &2.36\% &18.74\%\\
            & AR count & 1233 & 10 & 85 & 453 & 860 & 9 & 72 & 404 \\
            & AR total percentage & … & 0.81\% &6.89\% &36.74\% &69.75\% &0.73\% &5.84\% &32.77\%\\ \\
            
           Post sub-NOAA filtering & Images count & 132 476 & 271 & 2 467 & 14 965 & 87 298 & 286 & 2 786 & 24 403\\  
            & Images percentage & … & 0.2\% & 1.86\% & 11.3\% & 65.9\% & 0.22\% & 2.1\% & 18.42\%\\
            & AR count & 1148 & 9 & 77 & 352 & 793 & 7 & 48 & 336\\
            & AR total percentage & … & 0.78\% & 6.71\% & 35.19\% & 71.78\% & 0.7\% & 5.31\% & 31.71\%\\
        \hline
        \end{tabular}
    \end{table*}

    In this study, we use the image dataset compiled from data collected with the Helioseismic and Magnetic Imager \citep[HMI,][]{2012Schou} from the Solar Dynamics Observatory \citep[SDO,][]{2012Pesnell}. It is called the Spaceweather HMI Active Region Patch \citep[SHARP; ][]{Bobra2014} and it provides images of automatically tracked active regions in the form of patches. The SHARP dataset includes a variety of data such as line-of-sight magnetograms and remapped cylindrical equal area (CEA) images followed by several parameters for each patch. We train our models using line-of-sight magnetograms from SHARP, corresponding to the time interval from May 2010 to August 2021. It covers the end of the 24th solar cycle and the majority of the 25th. We couple the SHARP data with the GOES X-ray flare catalog to label the images by using their common identification number of active regions provided by the National Oceanic and Atmospheric Administration (NOAA). The GOES catalog is provided by the National Centers for Environmental Information (NCEI) and lists every solar flare observed since 1974 with their soft X-ray class, start time, end time, NOAA number, and many more parameters. We fetch it using the Heliophysics Events Knowledgebase (HEK) system through the Sunpy library \citep{sunpy_community2020}.
    \subsection{Data filtering}
    After removing SHARP files without a NOAA number (i.e, magnetic patches without a sunspot, and unlikely to produce any flare), 1\,579\,166 images remain in the dataset. If we used this whole dataset, a single epoch would take days to complete and the trainings of our models would last for months, especially with our limited resources. With this in mind, we decided to keep one image per hour. The SHARP dataset initially captures one image every 12 minutes. This selection reduces the dataset to 315\,374 images, which should be enough to produce a decent model without excessively long training.\\
    Our dataset at this point presents images captured by HMI with no conditional selection. Unfortunately, a significant part of the dataset is unusable or may negatively impact the training of the model due to the size and position of the image. If we kept every image from the whole disk, there would be a consequential difference between SHARP files due to the projection linked to the curvature of the Sun. As such, we removed every image of active regions without the whole image within 45\degr from the central meridian. In addition to impacting the prediction performance of our models, the size of the image can also limit their architecture. Through the convolution and pooling processes of a CNN, the image can become too small to fit in the architecture of our models or lose all its information before the flattening layer. Acknowledging this problem, we decided to keep SHARP files with a width and height greater than 150 pixels. This leaves us with 141\,378 images representing 1\,233 active regions.\\
    
    The dataset at this stage contains magnetograms cutouts (25417 images, 241 active regions) encompassing several smaller active regions (8902 images, 85 active regions). The metadata of those large cut-outs registers a list named NOAA\_ARS, enumerating the NOAA number of every region in the image. In the process of splitting the dataset into training, validation, and test dataset, there could be a larger active region with several smaller ARs in one dataset and magnetograms of the smaller regions in another one. If this is not handled properly, this could compromise the parsimony of the dataset. To deal with this issue, we decided to remove every SHARP file when they are sub-regions of a larger one.
    Through all this filtering, the dataset now contains 132476 images and 1148 active regions. 

    \subsection{Data labeling and dataset production}
    
    To have a better visualization and control of its distribution, we labeled each image with seven distinct image labels:
    
    \begin{enumerate}
        \item X: Images of an active region producing an X class flare in less than 24 hours.
        \item M: Images of an active region producing an M class flare in less than 24 hours.
        \item C: Images of an active region producing a C class flare in less than 24 hours.
        \item Future-X (FX): Images of an active region producing an X class flare in more than 24 hours.
        \item Future-M (FM): Images of an active region producing an M class flare in more than 24 hours.
        \item Future-C (FC): Images of an active region producing a C class flare in more than 24 hours.
        \item Never-Flare (NF): Images of an active region that will not produce any flare in the future.
    \end{enumerate}
    Images with several flares in the following 24 hours will be labeled using the strongest flare. Table \ref{table:1} presents the dataset distribution label-wise through every step of data filtering. \\
    For this study, we produced two different classifications. The first one is called CMX classification and includes C, M, and X images in the positive class, while the rest is in the negative class, and the second classification is called MX classification and only includes M and X images in the positive class. By training models on these two classifications, we could evaluate the difference in prediction performance for X and M classes between our classifications.\\
   \begin{figure*}
       \centering
   \includegraphics[width=\linewidth]{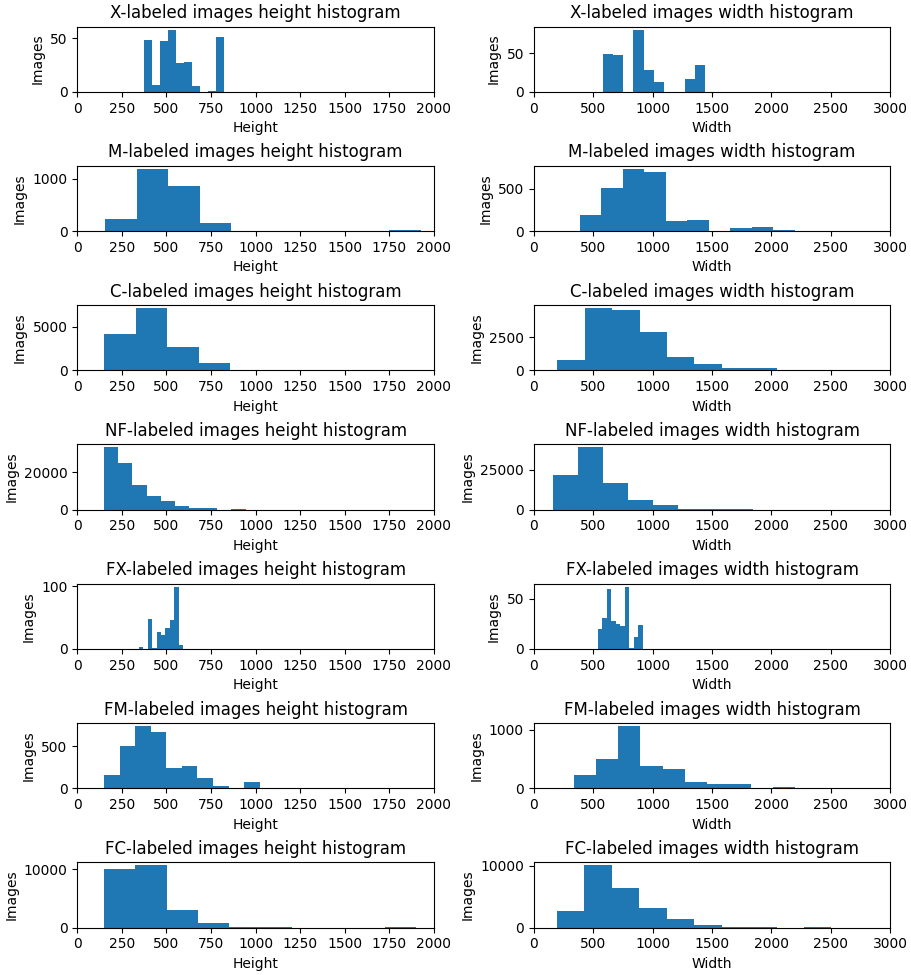}
   \caption{Width and height histogram of magnetograms per image label after filtering.}
              \label{FigHist}%
    \end{figure*}
    
    \noindent
    Furthermore, with this labeling system, we can produce the training, validation, and test datasets approximately similar to the real-life distribution inside each class. {The training datasets are used during the training of our models, while the validation datasets are used to ensure their training is progressing correctly. The test dataset is used at the end to evaluate the prediction ability of our models}. We first select the NOAA identification number of every active region with X-labeled images among the filtered dataset. We then shuffle this list and split them in an 8:1:1 distribution corresponding to the training, validation and test sub-dataset. This distribution is then used to store the images with the corresponding NOAA number to their respective sub-dataset. Following this, we remove these images from the filtered dataset and proceed with active regions represented by the remaining M-labeled files. We continue this process with the remaining labels. The process order is the following: X, M, C, NF, FX, FM and then FC. This operation ensures that there is never the same active region between the training, validation, and test dataset, keeping the parsimony of the dataset and guaranteeing a correct distribution of flares per class. \\    
    \begin{table*}
        \caption{Sub-dataset distribution after splitting and downsampling.}             
        \label{table:2}      
        \centering          
        \begin{tabular}{c c c c c c c c c c}     
        \hline\hline     
        Dataset & Classification & Quantity & X & M & C & NF & FX & FM & FC\\
        \hline                    
           Training & CMX & average number of image & 218 & 1918 & 11824 & 10665 & 35 & 344.0 & 2917\\
               & & Standard deviation & 13.35 & 81.36 & 154.79 & 205.05 & 9.32 & 27.81 & 90.19\\
               & & percentage of the dataset & 0.78\% & 6.87\% & 42.35\% & 38.19\% & 0.13\% & 1.23\% & 10.45\% \\\\
                    & MX & average number of image & 207 & 1955 & 370 & 2201 & 7 & 69 & 596\\
               & & Standard deviation & 12.1 & 62.45 & 18.03 & 50.88 & 2.9 & 11.03 & 27.76 \\
               & & percentage of the dataset & 3.84\% & 36.16\% & 6.84\% & 40.72\% & 0.13\% & 1.28\% & 11.02\%\\\\
            Validation & CMX & average number of image & 25 & 277 & 1578 & 8409 & 21 & 255 & 2467\\
               & & Standard deviation & 12.12 & 60.63 & 145.56 & 228.39 & 40.64 & 71.73 & 194.46\\
               & & percentage of the dataset & 0.2\% & 2.12\% & 12.11\% & 64.52\% & 0.17\% & 1.96\% & 18.93\% \\\\
                    & MX & average number of image & 37 & 240 & 1405 & 8773 & 23 & 315 & 2466\\
               & & Standard deviation & 15.58 & 37.56 & 165.69 & 322.65 & 38.83 & 129.88 & 353.82 \\
               & & percentage of the dataset & 0.28\% & 1.82\% & 10.6\% & 66.18\% & 0.18\% & 2.37\% & 18.58\%\\\\
            Test & CMX & average number of image & 26 & 271 & 1562 & 8949 & 27 & 264 & 2467\\
               & & Standard deviation & 12.51 & 60.41 & 145.08 & 393.78 & 29.12 & 88.07 & 288.46\\
               & & percentage of the dataset & 0.2\% & 2.01\% & 11.53\% & 65.94\% & 0.2\% & 1.95\% & 18.17\% \\\\
                    & MX & average number of image & 26 & 270 & 1489 & 8726 & 45 & 317 & 2533\\
               & & Standard deviation & 11.51 & 53.77 & 107.64 & 500.91 & 47.86 & 146.73 & 189.15 \\
               & & percentage of the dataset & 0.2\% & 2.02\% & 11.12\% & 65.04\% & 0.34\% & 2.39\% & 18.9\%\\
        \hline
        \end{tabular}
    \end{table*}
    
    \begin{table}
        \caption{Width and height statistics of magnetograms after filtering.}             
        \label{table:3}      
        \centering                          
        \begin{tabular}{c c c c c c c}        
            \hline\hline                 
            Feature & label & Mean & Median & Std\tablefootmark{a} & Min & Max \\
            \hline                    
            Height & X & 561.7  &  524.0  &  137.7  &  374  &  824\\
                   & M & 497.0  &  480.0  &  181.2  &  157  &  1926\\
                   & C & 427.6  &  392.0  &  180.3  &  150  &  1919\\
                   & NF & 292.4  &  258.0  &  123.3  &  150  &  944\\ \\
                   & FX & 503.8  &  525.0  &  57.8  &  345  &  594\\
                   & FM & 442.0  &  412.0  &  157.5  &  150  &  1023\\
                   & FC & 371.8  &  349.0  &  140.9  &  150  &  1902\\ \\
                    
            Width & X & 914.9  &  883.0  &  257.1  &  579  &  1444\\
                  & M & 914.7  &  896.0  &  287.1  &  386  &  2194\\
                  & C & 805.9  &  745.0  &  317.9  &  199  &  2505\\
                  & NF & 528.9  &  483.0  &  221.8  &  160  &  2266\\ \\
                  & FX & 707.4  &  696.0  &  98.0  &  539  &  917\\
                  & FM & 879.6  &  807.0  &  306.7  &  340  &  2201\\
                  & FC & 704.9  &  641.0  &  279.8  &  196  &  2505\\
        \hline
        \end{tabular}
        \tablefoottext{a}{Standard deviation}
    \end{table}

    \subsection{Class imbalance}
    
    By using this labeling system, we create a significant class imbalance due to the amount of X and M flares compared to C and Never-Flare images. Class imbalance is a major problem in DL and causes the models to prioritize the classification of the larger class and ignore the smaller one. Solutions to this problem are usually separated into two categories, called data augmentation and downsampling. Data augmentation is a method in computer vision to produce artificial training data by rotating, stretching, flipping, or using other means. This is usually a great way to tackle class imbalance while teaching models to recognize images in different forms. However, we cannot use this method to predict solar flares. Our images of active regions need to stay physically realistic based on our observation tools. Sunspots follow a logic such that by rotating an image for example, features such as the placement of negative and positive spots become unrealistic and if a model learns to recognize them, it may become confused during the testing phase and produce a lower prediction performance. However, downsampling is a possible solution to our problem. By randomly removing data from the majority class, we can balance the dataset at the expense of breaking the realistic flare distribution between classes. For the MX classification training dataset, we use downsampling with a 40:60 ratio, reducing the majority class data count to 1.5 times the quantity of data in the minority class. Although there is still a slight class imbalance, it is better than the previous 2:98 ratio. We decided not to lower the downsampling ratio anymore to keep a well-rounded distribution and enough data for the models to learn properly. If we chose to downsample to a 50:50 ratio, approximately 80 C-labeled images would have remained compared to 300 X-labeled images, which is not enough for a correct training. In the case of the CMX classification, we can use a 50:50 ratio due to the presence of the C-labeled images in the minority class, as it greatly increases the size of the dataset. If we used a greater ratio for the majority class, the training time would have increased with minimal improvement in the prediction performance of our models. The downsampling ratio for models trained using a Score-Oriented Loss function is adjusted to a 20:80 distribution as the SOL function helps to reduce the impact of the class imbalance issue and allows us to reduce the use of the downsampling method to use more data for our models training. Table \ref{table:2} presents the training, validation and test dataset distribution averaged across every cross-validation. Together with the downsampling method, we use a class-weight method, which increases the importance of correctly classifying the minority class for the model during its training, based on the data distribution between classes. This contributes to compensate for the class imbalance. The formula used to compute the class-weight for the c class is
    \begin{equation}
        Weight(c) = \frac{n\_samples}{n\_classes * n\_samples\_of\_class(c)}.
    \end{equation}
    
    \subsection{Data preprocessing}
    
    To fit the Trad-CNN input, we resize the magnetograms to 512×512 pixels using bi-linear interpolation through the TensorFlow resize function. 128×128, 100×100, and 198×198 are popular resize size choices. However, we chose this resize size to reduce the information loss in strong solar flares. Active regions with intense flares in the next 24 hours tend to be of a bigger size than low-activity active region (see Table \ref{table:3} and Fig. \ref{FigHist}). 
    If we resized them to a tenth or a fifth of their size, the magnetograms would lose a lot of their details and there are strong indications that flare productivity is related to small scales features (e.g., for features that can be observed in line-of-sight magnetograms: strong gradients around the polarity inversion line \citep{Jing_2006}, or the so-called “magnetic channel” \citep{Wang_2008_magnetic_channel} and “magnetic tongue” \citep{Poisson2016}).\\
    
    In the case of SPP-CNN models, there should be no need for resizing. However, to train our models, we use the mini-batch method. By feeding multiple images at a time to the model during its training and calculating the mean gradient to optimize its weight, the model can converge faster to its optimum state while offering more stable progress during its training. However, this method requires a common size for inputs among a mini-batch. To respect this condition while taking advantage of the SPP layer, we group images in 10 groups based on the sum of their width and height. Each group corresponds to an interval of 400 pixels. The first group contains images with the sum of their width and height higher than 300 pixels and less than 300+400=700 pixels. The interval of the second group is between 700 and 1100, and we continue for each group. During the training, mini-batchs are constructed by selecting images of the same group and then resized to the average size of the mini-batch. This process allows us to minimize the amount of resizing and therefore the loss of details. There was the possibility to feed our model images one by one but after many tests, we concluded that models delivered a better prediction performance with mini-batch while some models without this method were not able to learn at all. During our the trainings of our models, we used mini-batch size of 16 images.\\

    Magnetograms can present NaN values. This is either due to the image showing beyond the border of the Sun or artifacts appearing during its production. To avoid the spread of NaN values during the backpropagation, we replace them with zeros (0 is actually the average value of magnetograms, so this does not introduce much disturbance). Moreover, we normalize the input with a z-score normalization \citep{LeCun2012} from the SciPy library before feeding them to the CNN.


\section{Method}

    In the current state of computer vision, CNNs are a popular approach to recognize and classify images. Through multiple layers of data processing, they can automatically extract complex features from their inputs and then output a prediction by feeding them to fully connected layers. The main layers of CNNs \citep{Yamashita2018} are the following:
    
\begin{enumerate}

    \item Convolution layers are composed of input data, a convolutional window, and an output called a feature map. Their main goal is to extract features from their input by applying a convolution. This process can be described as adding each element of the image to its local neighbors weighted by the convolution window. A convolution layer is illustrated by the number of filters, a filter size, a stride number, and a padding mode. 
    \item Activation layers are usually used to apply nonlinear functions, such as rectified linear unit \citep[ReLU][]{agarap2019deep}, softmax and sigmoid, to their inputs. 
    \item Pooling layers are built the same way as convolution layers except that their goal is to extract the maximum value, or average depending on the use, in the pooling window. They are designed to reduce the input dimensionality while keeping the most important features.
    \item A single flattening layer is used to transform its input, which is usually feature maps, from an n-dimensional matrix to a one-dimensional array. This allows the data to fit in the input of the first fully connected layer.
    \item Fully connected layers are the last step of a convolutional neural network. They are composed of neurons and each neuron of the current layer is connected to every neuron of the previous layer with weights between them. The last fully connected layer is composed of n neurons, where n is the number of classes in case of a number of class greater than two. For a binary classification, using a single output neuron associated with a threshold value between the two classes is enough to classify the data.

\end{enumerate}

    \begin{table*}
        \caption{SPP-CNN and Trad-CNN hyperparameters.}
        \label{table:4}
        \centering
        \begin{tabular}{c c c c c c c c} 
            \hline\hline 
                Classification & Architecture & Loss & Learning Rate & Mini-batch Size & Resizing Size & Downsampling Ratio & Epoch \\
            \hline     
                CMX & Trad-CNN & BCE & $1.0 x 10^{-6}$ & 16 & 512×512  & 50:50 & 60\\
                    &          & TSS & $5.0 x 10^{-6}$ & 16 & 512×512  & 20:80 & 60\\
                    & SPP-CNN  & BCE & $1.0 x 10^{-5}$  & 16 & …       & 50:50   & 60\\
                    &          & TSS & $5.0 x 10^{-6}$  & 16 & …       & 20:80   & 60\\
            \hline
                MX  & Trad-CNN & BCE & $2.0 x 10^{-6}$ & 16 & 512×512  & 40:60 & 100\\
                    &          & TSS & $1.0 x 10^{-6}$ & 16 & 512×512  & 20:80 & 100\\
                    & SPP-CNN  & BCE & $1.0 x 10^{-5}$  & 16 & …       & 40:60   & 100\\
                    &          & TSS & $1.0 x 10^{-5}$  & 16 & …       & 20:80   & 100\\
        \hline
        \end{tabular}
    \end{table*}
    A traditional CNN is composed of a sequence of convolution layers, activation layers, and pooling layers. The end of the network is constituted of fully connected layers. The transition from one layer to another is based on parameters such as weights and bias and can amount up to billions depending on the size of the network. These parameters are fine-tuned by learning and adapting to the data by using gradient-based optimization and other methods. This is an important step to teach a neural network to classify an image, and it is referred to as training. The training of our models was composed of three main processes:
\begin{enumerate}
    \item The first step is called forward propagation. We first feed pre-labeled data from a training dataset to the model one by one. It will try to predict their class and based on the output of the network, an error value will be calculated with an error function. Error functions, or loss functions, are chosen based on the type of data, the classification method, and many other parameters. Popular error functions include binary and categorical cross-entropy and mean squared error (MSE).
    \item The second step is called backpropagation. We fine-tuned the weights of our various layers by using the error value. We use the chain gradient rule to calculate the gradient of each weight in the network from the last layer to the first, and then update them. 
    \item After the whole training dataset was fed to the model through forward propagation and backpropagation, we used the validation dataset to track the prediction performance of our model and the training efficiency. This also allows the automatic adjustment of some hyperparameters \citep{kingma2017adam, you2019does} of the model for an optimal training. If there is no error and the model keeps learning, we continue the training of our model by shuffling the training dataset and repeat the previous process several times.
\end{enumerate}
   \begin{figure*}
       \centering
   \includegraphics[width=\linewidth]{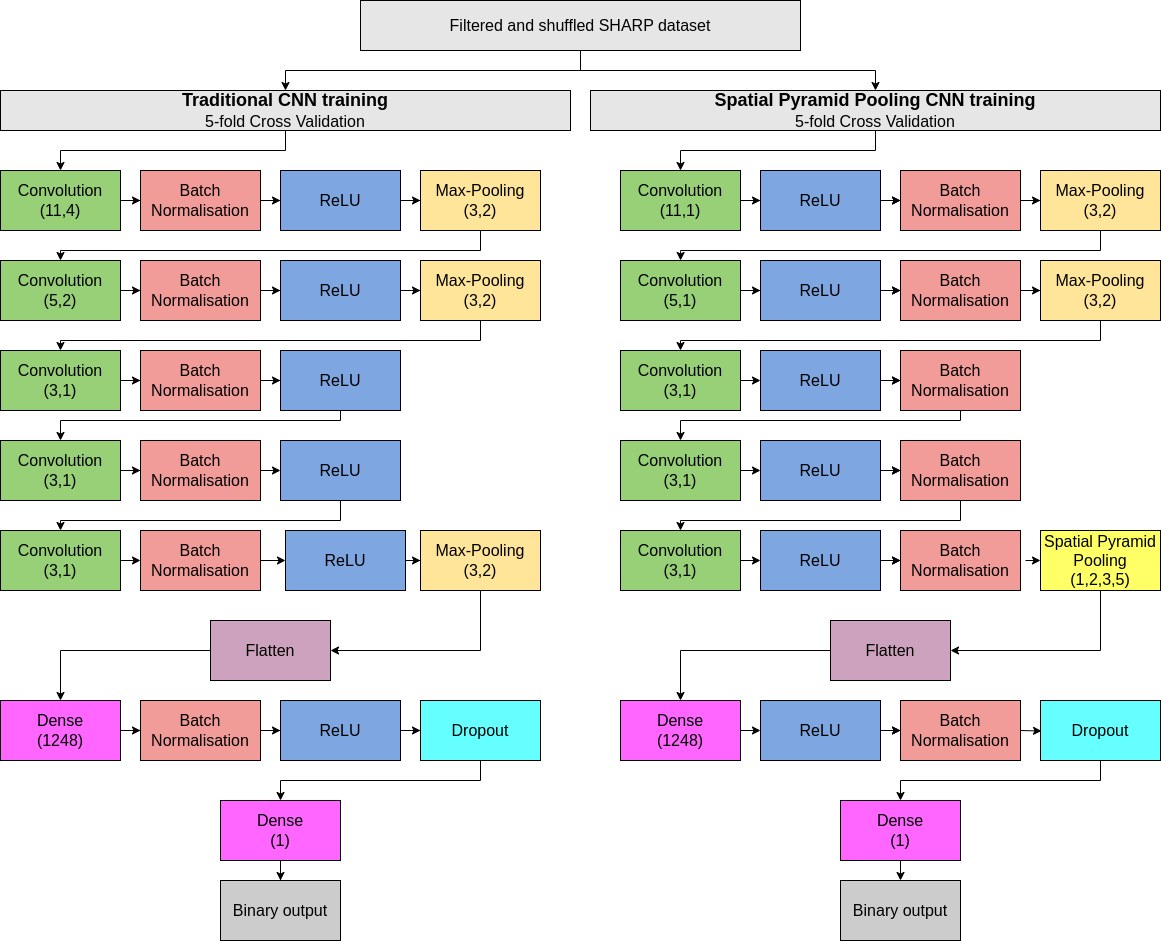}
   \caption{Diagram of Trad-CNN and SPP-CNN architectures.}
              \label{ModelDiagram}
    \end{figure*}
    Each cycle of these three processes is called an epoch. Whenever a fixed number of epochs is achieved or when the metrics used to evaluate the efficiency of the prediction of our models stagnate, the training is stopped. This set of epochs forms one fold of a k-fold cross-validation. For each fold, we keep the best model produced among all epoch by selecting the model with the highest sum of validation precision and recall. The precision and recall are metrics used to evaluate the prediction performance of a model, and they will be explained at the beginning of section 4. This method allows us to select the best model to predict solar flares accurately and without over predicting them. Following this, we evaluate the selected models with their test datasets to ascertain their true prediction performance.\\
    
    As mentioned, a CNN has several hyperparameters that need to be fixed beforehand or adjusted during the training such as the learning rate, the momentum, the loss function and the mini-batch size. These hyperparameters are crucial to train correctly a model as they can significantly impact its learning process. For instance, with a learning rate too low, the model may learn too slowly or never converge. However, the model may never reach its optimal state if the learning rate is too high. Another variable aspect that can have a significant impact on the prediction ability is the architecture of the model. There is an infinite combination of layout and layers and the data can become over-transformed, unusable, or under-processed and too complex if the wrong architecture is used. However, with the right architecture, one can achieve unmatched predictive performances \citep[see][]{alexnet}.\\
    
    During this study, we have trained and analyzed several model architectures before choosing which one to use for this paper. Each had varying layer composition and hyperparameters, such as implementing or removing batch normalization layer \citep{ioffe2015batch} and dropout \citep{JMLR:v15:srivastava14a}. We fine-tuned the learning rate, the mini-batch size, convolutional layers feature window strides and sizes, number of layers and many more hyperparameters and elements of the architecture to obtain models with significant flare prediction performance. The differences between all architectures were sometimes minimal, but each modification brought a lot of variation in the prediction performances of our models. The model architectures and hyperparameters used in this paper were chosen based on the average prediction performance and stability across a 5-fold cross-validation.\\
    
    In this paper, we analyze and discuss the result of two model architectures that stood out while searching for the best architecture for our study. The notation we use to describe deep neural networks is based on the one used by \citet{ciresan2012multicolumn}. To illustrate the notation, we provide the following description:
    \begin{align*}
    \begin{split}
    &16x512x512-32C(11,4)-BN-A(ReLU)-MP(3,2)-\\
    &32C(5,2)-BN-A(ReLU)-SPP(1,2,3,5)-300FC-\\
    &BN-A(ReLU)-DO(0.5)-1FC-A(sigmoid).
    \end{split}
    \end{align*}
It represents a network with 16 input images of dimension 512×512; a 32-feature convolution layer with an 11×11 square feature window and a stride of four; a batch-normalization layer; an activation layer with a ReLU function; a max pooling layer with a 3×3 window and a stride of two; a 32-feature convolution layer with a 5×5 feature window and a stride of two; a batch-normalization layer; an activation layer with a ReLU function; an SPP layer with splits of size one, two, three, and five; a fully connected layer with 300 hidden units; a batch-normalization layer; an activation layer with a ReLU function; a dropout layer with 50\% probability; a fully connected layer of one output neuron; and an activation layer with a sigmoid function.\\
    The first architecture we analyzed is referred to as Trad-CNN and is described as follows:
    \begin{align*}
    \begin{split}
    &16x512x512-32C(11,4)-BN-A(ReLU)-MP(3,2)-\\
    &32C(5,2)-BN-A(ReLu)-MP(3,2)-32C(3,1)-BN-\\
    &A(ReLu)-32C(3,1)-BN-A(ReLu)-32C(3,1)-BN-\\
    &A(ReLu)-MP(3,2)-1248N-BN-A(ReLU)-DO(0.5)-\\
    &1N-A(sigmoid).
    \end{split}
    \end{align*}
    The second architecture, SPP-CNN, is the same as Trad-CNN but swaps the position of the batch-normalization layers with the activation layers. In addition, it sets the stride of convolutional layers to one and implements an SPP layer \citep{He2014_spp} before the flattening layer instead of a max pooling layer. A spatial pyramid pooling layer is a layer that slices its input into a fixed number of tiles based on its parameter and selects the maximum value in each tile. This process produces a fixed input for the next layer and is especially interesting before the flattening layer, as fully connected layers are the only ones that need a fixed input. The SPP layer allows us to input images of any size in the models and to retain all their details otherwise lost due to the resizing process.
    The architecture of the SPP-CNN models is described as the following: 
        \begin{table*}
        \caption{Average results of evaluation of cross-validation of SPP-CNN and Trad-CNN based on classification and loss function.}
        \label{table:5}
        \centering
            \begin{tabular}{c c c c c c c c c c c c c} 
            \hline\hline 
                Classification & Architecture & Loss & TP & TN & FP & FN & Accuracy & Precision & Recall & TSS & TSS Std\tablefootmark{a} & PR AUC \\
            \hline     
                CMX & Trad-CNN  & BCE & 1310 & 9584 & 2150 & 545    & 0.8 & 0.38 & 0.7 & 0.52 & \underline{0.02} & 0.52\\
                    &           & TSS & 1369 & 9713 & 2118 & 502    & 0.81 & 0.39 & 0.73 & 0.55 & 0.06 & 0.57\\
                    & SPP-CNN   & BCE & 1401 & 10429 & 1253 & 464   & \underline{0.87} & \underline{0.56} & \underline{0.76} & \underline{0.65} & 0.16 & \underline{0.68} \\
                    &           & TSS & 1374 & 7834 & 3691 & 497    & 0.69 & 0.31 &    0.74 & 0.42 & 0.06 & 0.43 \\
            \hline
                MX  & Trad-CNN & BCE & 242 & 7622   & 5779 & 71     & 0.58 & 0.06 & \underline{0.77} & 0.35 & 0.17 & 0.08 \\
                    &          & TSS & 177 & 10884  & 2323 & 93     & 0.82 & 0.08 & 0.67 & \underline{0.5} & \underline{0.09} & 0.11 \\
                    & SPP-CNN  & BCE & 174 & 10827  & 1994 & 107    & \underline{0.84} & 0.1 & 0.62 & 0.46 & 0.24 & 0.14\\
                    &          & TSS & 217 & 9460   & 3614 & 81     & 0.73 & \underline{0.11} & 0.7 & 0.43 & 0.28 &\underline{0.2}\\
        \hline
        \end{tabular}
        \tablefoottext{a}{True skill statistic standard deviation}
        \tablefoot{Underlined values highlight the best result between every models using the same classification.}
        
    \end{table*}
    \begin{align*}
    \begin{split}
    &input-32C(11,1)-A(ReLU)-BN-MP(3,2)-32C(5,1)- \\
    &A(ReLu)-BN-MP(3,2)-32C(3,1)-A(ReLu)-BN-\\
    &32C(3,1)-A(ReLu)-BN-32C(3,1)-A(ReLu)-BN-\\
    &SPP(1,2,3,5)-1248N-A(ReLU)-BN-DO(0.5)-\\
    &1N-A(sigmoid).
    \end{split}
    \end{align*}
    The SPP layer allows us to use inputs of different size, which is why there is no input size specified in the description of SPP-CNN. However, this advantage also brings a problem in the form of the minimum size of the input. As previously mentioned, we set the stride of our convolution layers to 1, which is to prevent images from disappearing. If we used the same parameters as the Trad-CNN architecture, we would be forced to exclude images smaller than 300 pixels in either width or height due to the data size reduction induced by the convolution and max pooling layers. The output of these layers was calculated with the following equation: 

    \begin{equation}
        Output\_edge\_size = \frac{W-K+2P}{S}+1,
    \end{equation}
    
    where W is the width or height of the input), K the window size, P the padding and S the stride. With a convolution window stride of 1, we can restrict the exclusion of images to images with edge size lower than 150 pixels in width or height. This exclusion was done in datasets of both model architectures, during the data filtering process.\\
    Another difference between the Trad-CNN and SPP-CNN architectures is the swap of order of the batch-normalization and activation layers. We found that the models based on the SPP-CNN architecture cannot be trained effectively if their batch-normalization layers are placed prior to the activation layers. The models ended up predicting every input as either zero or one. The exact reason is unknown and deserves further studies in future papers.\\
    
    In this study, we also compare the results of our models trained using two distinct loss function : the Binary Cross-Entropy (BCE) and a Score-Oriented Loss \citep[SOL;][]{2022Marchetti_SOL} function optimizing the True Skill Statistics \citep[TSS;][]{Bloomfield_2012} metric. The BCE is a widely used loss function while the SOL, aiming to optimize a given score (or metric), is not available in the Keras libraries in opposition to the BCE. In this study, we implemented the SOL function manually while not knowing that it is available on GitHub\footnote{\url{https://github.com/cesc14/SOL}}. In our case, the TSS is an interesting score to optimize due to its insensitivity to class imbalance. As the SOL function helps reduce the impact of the class imbalance problem, we trained our models, using this loss function, with a downsampling ratio lower than when using a BCE loss function. Reducing the downsampling ratio while using other class imbalance solutions increase the diversity of our dataset and can help improve the prediction performance of our models.\\

    The architecture of our models is represented in the Fig. \ref{ModelDiagram} and the various hyperparameters used for the trainings of each model are specified in Table \ref{table:4}. The optimizer used during trainings is Adam, and its parameters are the default ones used in the Keras library. Every convolution in both architectures use the padding method “same.”\footnote{\url{https://www.tensorflow.org/api_docs/python/tf/keras/layers/Conv2D}}\\
    The models of this study were developed mostly using Python with the NumPy, Pandas and TensorFlow libraries and processes were run using an A100 graphical processing unit.
    

\section{Result}

   \begin{figure*}
       \centering
       \includegraphics[width=\linewidth]{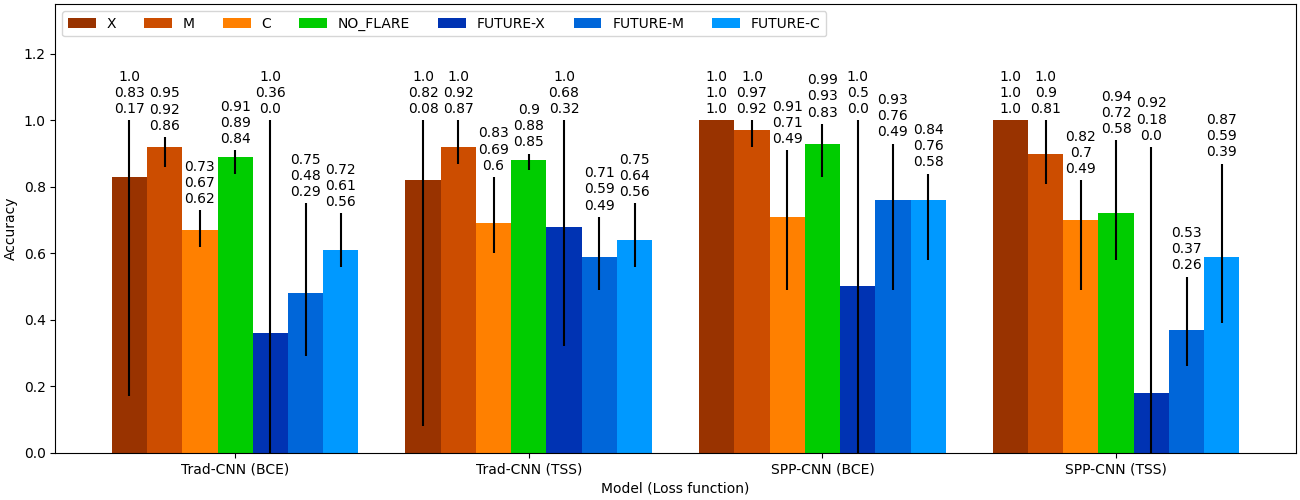}
       \caption{Prediction accuracy of models trained on the CMX, as distributed between each label. The error bars on each bar of the chart represent the range of accuracy between the best and worst model for each label. The three numbers above each bar represents, from top to bottom, the maximum accuracy, the average accuracy, and the minimum accuracy for each label.}
        \label{Accuracy_per_label_CMX}
    \end{figure*}
   \begin{figure*}
       \centering
       \includegraphics[width=\linewidth]{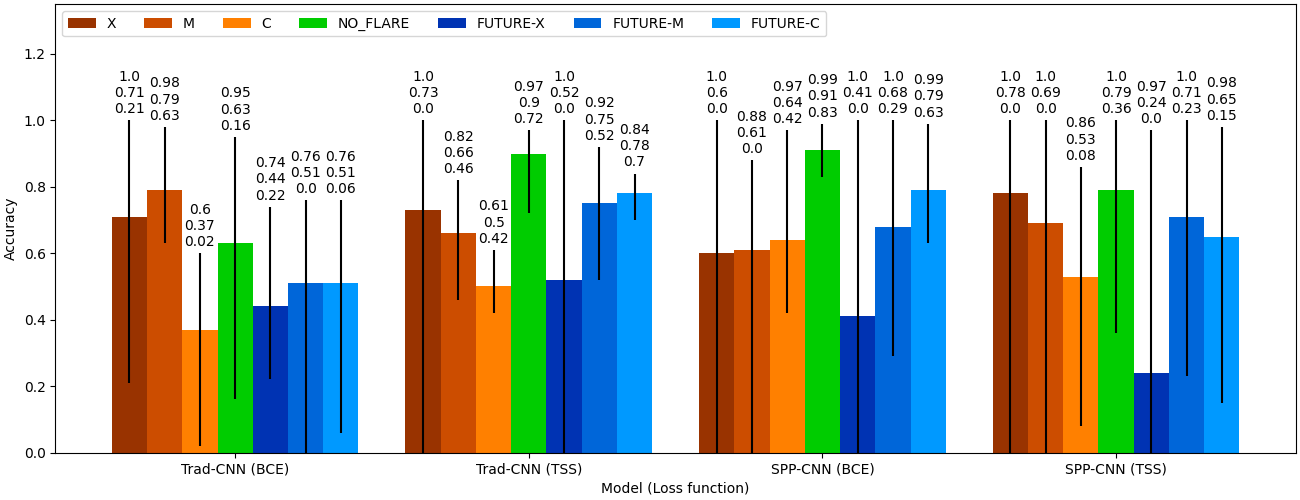}
       \caption{Same as Figure \ref{Accuracy_per_label_CMX} but for the MX classification.}
        \label{Accuracy_per_label_MX}
    \end{figure*}
    
    \subsection{Evaluation metrics}
    
    The models of this study output a binary classification. For the CMX classification, an image is classified as positive if it presents an active region that is flaring in less than 24 hours, or considered negative in any other case. In contrast, the MX classification only considers X and M labeled images as positive. Based on the label of an image and its prediction by the model, we can deduce four possible outcomes. A prediction can be:
    \begin{enumerate}
        \item A true positive (TP), which means the prediction of the model and the label of the image are positive
        \item A true negative (TN), which means the prediction of the model and the label of the image are negative
        \item A false positive (FP), which means the prediction of the model is positive, but the label of the image is negative
        \item A false negative (FN), which means the prediction of the model is negative, but the label of the image is positive
    \end{enumerate}
    With these four outcomes, we can evaluate a model by making it predict the label of every image from a test dataset and build the confusion matrix which enumerates how much TP, TN, FP, and FN were output. From this matrix, it is then possible to calculate the characteristics of our models. We use :

    \begin{enumerate}
        \item The recall, which shows the proportion of images with a positive label correctly predicted: 
    \begin{equation}
        Recall = \frac{TP}{TP+FN}.\\
    \end{equation}
        \item The precision, which shows the proportion of images predicted positive with a positive label: 
    \begin{equation}
        Precision = \frac{TP}{TP+FP}.\\
    \end{equation}
        \item The accuracy, which shows the proportion of correct prediction: 
    \begin{equation}
        Accuracy = \frac{TP+TN}{TP+TN+FP+FN}.\\
    \end{equation}
    \end{enumerate}
    
    As much as these three characteristics present useful information, they are prone to class imbalance. Therefore, we focus our analysis on the Precision-Recall Area Under Curve (PR AUC) and the True Skill Statistics, which are less sensitive to this problem and more relevant to our case. The PR AUC is produced by computing the Precision-Recall Curve and evaluating the area under this curve \citep{https://doi.org/10.1111/2041-210X.13140} while the TSS is calculated by the following formula:
    \begin{equation}
        TSS = \frac{TP}{TP+FN} - \frac{FP}{FP+TN}.
    \end{equation}

    The trainings of our models were split into several epochs during which each model was trained on the whole training dataset and then evaluated on the validation dataset. Throughout these epochs, the validation accuracy, recall, and other metrics of our models varied based on how much the models were underfitting or overfitting. At the end of each training, the model with the best sum of validation-recall and validation-precision among the different epochs was kept and then evaluated using the corresponding test dataset. 
    
    \subsection{Results}
    
    The result of those evaluations are shown in Table \ref{table:5}. These values were calculated by averaging {the result of the evaluation of each group of models} across their cross-validation. At a first glance, the Trad-CNN models in the CMX classification section present equivalent results no matter the loss function, whereas the results of the SPP-CNN models indicate them to be better at predicting flares when trained using the BCE loss function on the CMX classification. The results indicate an improvement of 0.25 in PR AUC, 0.23 in TSS and 0.25 in precision. Moreover, the evaluation results of the SPP-CNN models trained using the BCE loss function are on average higher on every metric than the results of the Trad-CNN models trained with the TSS-based loss function. We notice in this case an improvement of 0.1 in the TSS metric, 0.17 in precision and a 0.11 improvement in PR AUC metric. Given the standard deviations of the TSS of our models, we cannot conclude one model to be strictly superior to another. However, using the Kolmogorov-Smirnov test\footnote{\url{https://docs.scipy.org/doc/scipy/reference/generated/scipy.stats.kstest.html}} on the TSS metric obtained throughout the cross-validation folds of our models, we acquire a confirmation at 92\% chance for the SPP-CNN models trained with the BCE loss to follow a different probability distribution than SPP-CNN models trained with the SOL function. This allows us to infer that there is a 92\% chance that using BCE to train SPP-CNN models was better than the TSS-based loss. On the other hand, the results of the evaluations of our models using the MX classification are more nuanced, there is no superior model. Furthermore, although the models achieve promising scores in recall and TSS, the best precision observed is 0.11 for the TSS-based loss trained SPP-CNN model, which is too low to consider using them for operational purposes or any explainable artificial intelligence (XAI) analysis.\\

    While these statistics and indices highlight the overall prediction performance of our models, they do not give any information on the type of flare our models are able to predict. Using a CMX classification, our model could have a recall of 0.8 while correctly predicting only C class flares. To acquire a better understanding of the capability of our models, we analyze their predictions results on the test dataset for each image label and the evolution of the prediction confidence of a flare, i.e. the output value of the last neuron of our models, during the entire lifetime of a given active region.
    Table \ref{table:6} and \ref{table:7} present the number of correct predictions, the accuracy and the standard deviation of the accuracy, based on the label, the model, the loss function and the classification. Figures \ref{Accuracy_per_label_CMX} and \ref{Accuracy_per_label_MX} illustrate the best, the worst, and the average prediction accuracy per label for every model trained. These tables and figures indicate an improvement in prediction performance of SPP-CNN models compared to Trad-CNN models for the CMX classification. The first result we can note is the perfect prediction accuracy of the SPP-CNN models for images linked to X class flares and very high accuracy for images linked to M class flares. The second result is the very low accuracy for FX-labeled images compared to other images.\\
    
    Figures \ref{pred_conf_SPP_BCE_CMX_X} to \ref{pred_conf_TRAD_TSS_CMX_M} in the appendix highlight the evolution of the prediction confidence of the models trained using the CMX classification, for the prediction of every image of a single active region. Figures \ref{pred_conf_SPP_BCE_CMX_X} to \ref{pred_conf_TRAD_TSS_CMX_X} present the evolution of the prediction confidence when predicting images from ARs producing X class flares, whereas figures \ref{pred_conf_SPP_BCE_CMX_M} to \ref{pred_conf_TRAD_TSS_CMX_M} present the evolution of the prediction confidence for images from ARs producing an M class flare as the strongest flare of their lifetime. Starting from the first image of each active region within the test dataset, the x-axis represents the time in hours, with markers illustrating the occurrences of solar flares produced by the corresponding active region. The y-axis represents the prediction confidence of the corresponding model, and the horizontal line at 0.5 represents the threshold that separates a positive prediction (1) from a negative prediction (0). To ease the analysis of the figures, parts of the chart with the colored area correspond to the interval of time during which images are labeled positive in our test dataset.

\section{Discussion}
    Our results in terms of accuracy, precision and especially TSS are in line with previous studies \citep[see for example the review in Table 2 of][]{Li2020}. An exact comparison, however, is difficult because of the diversity in the data selection and labeling methods, as presented in Section 1. This is why we focus here on comparing the results of our two model architectures, that were trained on the same images dataset and labeling systems. Furthermore, while the results such as recall and label-wise accuracy vary greatly through our models and classifications, we can analyze their causes in the light of the methodology we used.
    
    \subsection{Comparison of the results between architectures and loss functions}
    
    The first noteworthy point in our results is the relative improvement of the SPP-CNN compared to the Trad-CNN to predict flares using the CMX classification. There is indication that associating the SPP layer with the BCE loss function increase the rate of successful prediction of flares while reducing the rate of false alarm compared to every other couple of architectures and loss functions. An average increase of 0.1 in the TSS metric compared to Trad-CNN models using the TSS-based loss and a perfect classification of X-labeled images is especially notable.\\
    While SPP-CNN models trained with the BCE loss function present impressive results, we can observe a drop in our metrics when using the Score-Oriented-Loss function instead of the BCE loss function. We speculate the SPP layer to be incompatible with the more deterministic predictions offered by the SOL function compared to the BCE loss function, which can be observed in the figures \ref{pred_conf_SPP_BCE_CMX_X} to \ref{pred_conf_TRAD_TSS_CMX_M}. The process of SPP consist in slicing its input (i.e, feature maps) in tiles and selecting the maximum value in each of those tiles. This means extracting the presence and intensity of a feature within a tile while factoring out the quantity of such feature. This process transforms the output of each tile to either “presence” or “non-presence” of the feature along with its intensity and can overlap with the deterministic prediction associated with the SOL function. \\
    
    While the prediction ability of SPP-CNN models decrease when using a SOL function, Trad-CNN models present a slight improvement in prediction ability when trained with it compared to the BCE loss function. Although the gap of TSS is more noticeable for the MX classification, the difference is so small we cannot deduce whether the use of a SOL is the source of such improvement or if it originates from the random splits of our dataset. We would need further cross-validation fold to obtain a reliable average. Nevertheless, our results show potential in the use of a SOL function for flare prediction using more traditional architectures of models, although the prediction confidences in Figures \ref{pred_conf_TRAD_TSS_CMX_X} and \ref{pred_conf_TRAD_TSS_CMX_M} look more uncertain through the numerous sudden swap in prediction during the evolution of a single active region. \\
    
    Aside from the difference linked to the impact of the loss function between our architectures, we can observe the stability of the training of our SPP-CNN models in the Figure \ref{Accuracy_per_label_CMX}. The classification accuracy of images of active regions leading to stronger flares is especially stable compared to active regions producing weaker flares or not producing any flare at all. The accuracy of the SPP-CNN models also decreases with the strength of observed flares whereas Trad-CNN models classify more accurately images of active regions producing flares of medium to low intensity or not producing any flare at all compared to strong, X flares.
    
    \subsection{Comparison of the results between classifications systems}
    
    The second point to note is the disparity in recall, TSS and especially precision and PR AUC throughout the results of our models between the CMX and MX classifications, no matter the loss function. The average (Recall, TSS, Precision, PR AUC) values for all models using the CMX classification are (0.73,0.53,0.41,0.55) while they become (0.69,0.43,0.09,0.13) for models using the MX classification.\\
    
    The results for the CMX classification demonstrate the excellent ability of our models to predict flares in the next 24 hours, especially with the SPP-CNN architecture. However, they are unable to output any reliable positive prediction using the MX classification. The average precision of 0.09 highlight the inability to correctly differentiate a negatively labeled image from a positive one. Furthermore, the disparity of prediction ability of our models between the CMX and the MX classifications are clearly visible while comparing the Figures \ref{Accuracy_per_label_CMX} and \ref{Accuracy_per_label_MX}. The accuracy and error bars of the Figure \ref{Accuracy_per_label_MX} show an instability of prediction performance among the models of a same classification and loss function, and their inability to correctly classify images between positive (flare) and negative (non-flare). It would be interesting to obtain a deeper view of the functioning of our models and where their attention lies. However, there are no “off-the-shelf” Explainable Artificial Intelligence tools available that would help in visualizing what attracts the attention of the models in the images when an SPP layer is used. Developing such tools may be the subject of a future study. Nevertheless, we can analyze and speculate on the causes of our results.\\
    
    The gap between the results of our models trained on the CMX classification and the MX classification can be explained by two reasons: the extreme class imbalance in the datasets of the MX classification and the increasing difficulty in differentiating flares for the MX classification due to the presence of C-labeled images in the negative class. \\
    Firstly, including C-labeled images in the negative class lead us to a dataset distribution of 2\% of positive-classified images to 98\% of negative-classified images  (to be compared to 13\% and 87\%, respectively, in the case of the CMX classification). Although several efforts were made to reduce the class imbalance in this study such as using class-weights, downsampling and Score-Oriented-Loss function, the amount of positive-class images is too low (2000 images used for training and 200 for the test dataset) to properly train our models.\\
    Secondly, the presence of C-labeled images in the negative class raises three major problems : 
    
    \begin{enumerate}
        \item The first problem is linked to the presence of C class flares in active regions producing M or X flares. While the triggering mechanisms of a C class flare may be different from an M or X flare, an active region producing a C flare followed by a higher class flare a few hours later may not present a significant evolution enough to be seen in a line-of-sight magnetograms. The similarity of images from such active regions can confuse our models in the way that two magnetograms observed within a 1-hour interval are supposed to be classified as positive and negative, respectively, without significant differences.
        \item The second problem raised is linked to the classification used in the GOES catalog. 
        The five classes of the GOES classification (A, B, C, M and X) follow a logarithmic scale representing the intensity of soft X-ray emission of a flare and each flare class is composed of subclasses from 1 to 9. This is useful when discussing flares, but when it comes to training a CNN, it becomes harmful to our prediction performance. This is because, in the GOES classification, a C9.0 flare will be classified in another class than an M1.0 flare while not presenting a significant difference. This is also the case between M9.0 and X1.0 flares and between B9.0 (which is labeled as an NF image in our labeling system) and C1.0 flares. Labeling our images based on this classification model can be confusing for the DL model as two images without significant differences can be in separate GOES classes and therefore in separate data classes. Furthermore, this issue can explain the ability of our models to correctly predict stronger flares better than weaker ones in the positive class. It is harder to differentiate images of flares close to the threshold separating flaring and non-flaring images. 
        \item The third problem can explain the difference of prediction performance between CMX and MX classifications for models following the SPP-CNN architecture:
        some features leading to solar flares may appear in an active region with an incoming flare of class C, M, and X flares. The SPP can efficiently detect the presence and intensity of such feature. However, one of the differences between C, M, and X-labeled line-of-sight magnetograms may be its quantity. While they may be distinguishable in a regular CNN, it is hardly possible with SPP-CNN models. As mentioned before, this is because the SPP layer slices its input into a fixed amount of tiles before selecting the maximum value in each tile, extracting the presence and intensity of a feature while factoring out the amount of time it appeared. If we were to input an X-labeled image with a very complex structure and a feature that appears multiple times, it would produce the same output as if {the feature} had appeared only once in a simple structure from a C-labeled image. This can be interpreted as if our SPP-CNN models can process features inherent to the triggering of solar flares, no matter their GOES class. However, their prediction abilities may be solely based on these features and whenever we mix C-labeled images in the negative class, it becomes ineffectual. In other words, the models may be able to predict flares with precision, but they do not differentiate between their class at all. This is the drawback of the coarse slicing used in the SPP layer and can be bypassed by increasing its parameters, which represents the number of tiles the input has to be sliced into. However, this would also drastically increase the computational power needed to train our models and to run a prediction. 
    \end{enumerate}
    The drop of predictive performance from models trained on the CMX classification to models trained on the MX classification is also observed in other studies, such as \cite{Li2020}.
    
    \subsection{Analysis of results label-wise}
    A third noteworthy point is the significantly lower prediction accuracy of FX-labeled images compared to all other labels. This may be due to the large range of time intervals between the manifestation of flare triggering mechanisms and the actual flare. An active region may keep a complex magnetic configuration during several days before triggering a flare, trigger a flare a few hours after the manifestation of flare triggering mechanisms, or present important signs of an incoming eruption while not erupting at all. We can observe the impact of this phenomenon in the Figures \ref{pred_conf_SPP_BCE_CMX_X} to \ref{pred_conf_TRAD_TSS_CMX_M}, especially in the chart of the evolution of the prediction confidence of the ''Cross\_validation number 1'' in the Figure \ref{pred_conf_SPP_BCE_CMX_M} where the prediction confidence remains close to 1 during several days after the last eruption and in the ''Cross\_validation number 3'' chart of the figure \ref{pred_conf_SPP_BCE_CMX_X} where the prediction confidence increase only 16 hours before the first flare. In these conditions, labeling our data based on a fixed prediction window separating images of flaring (positive) and non-flaring (negative) active regions can greatly confuse our models and impede on their prediction ability.
    
    \subsection{Discussion on the methodology and data used}
    
    With this study, we can assert the strengths and weaknesses of the SPP layer through the results of our models. However, there were various drawbacks to our method, data, and labeling system. One of the major problems is the class imbalance, reflected through the low number of active regions with an X class flare. Due to the projection problem with line-of-sight magnetograms, we needed to remove every image beyond 45 degrees from the central meridian of the Sun. Unfortunately, this process removed 605 images representing 17 active regions producing X class flares. We also discarded SHARP images based on their active region also being present in other bigger files and several other data modification and selection processes during our study. After the selection and preprocessing phase, only nine distinct active regions producing X class flares remained. These pre-selection methods were not finely tuned to obtain the optimal result while keeping the most images possible. They were established on the average protocol found during our bibliographical survey. Studying which selection and preprocessing method is optimal and how they impact the training is a process we could work on in the future. Furthermore, this could help us establish a way to reliably compare studies using DL for solar flare forecasting but with different methods of handling data. \\
    
    Although we obtained satisfying prediction skill scores for models using the SPP-CNN architecture, we believe further improvements can be made. A major limiting factor of this study is that we had to match as much as possible the architectures of the SPP-CNN model to that of the more traditional architectures, to compare their results and to prove the possible improvements brought by an SPP layer. This may have limited its predictive ability. Now that this comparison has been made, we can work on optimizing the SPP architecture and compare its performances with the SPP architecture of the present study.\\
    The use of the Score-Oriented-Loss function and other loss functions can also to be studied for the specific purpose of flare prediction and to improve the prediction performance of DL models as we saw a relative improvement in the prediction ability of our Trad-CNN in predicting flares with a SOL function compared to the use of a BCE loss function.\\
    Another improvement possible in this study is the labeling system for wider uses and better predictions. We used images with a flare in less than 24 hours as the positive class. However, there may be very little evolution in this time range, and it is reflected in the difference in accuracy for X-labeled images and FX-labeled images for the CMX classification. While the classification accuracy for X-labeled images is nearing a 100\% success rate, the FX-labeled image classification rate is close to a random classification. For future studies, we could choose 48h or even 72h to label flaring images. The best solution would be to train another neural network or implementing another output in our CNN, dedicated to predict the time before the eruption. Furthermore, the use of the GOES classification to label our data raised several problems. A better way to label our images or to predict flares would be based on the intensity in soft X-ray emission of each flare. In this manner, the model would not be confused anymore about which class an image belongs to. This would change the subject from a classification problem to a regression.\\
    While line-of-sight magnetograms are easy to use for DL, they are also limited in their information. The main features our models could learn about are the shape of the polarity inversion line and the size of active regions present in the image. This limits the ability of our modes to predict flares. By using other data types (like vector magnetograms), we may be able to extract more information about the solar flare-triggering mechanisms. Apart from changing our data source, we can also study the use of other DL methods. Convolutional neural networks are easy to use and efficient in recognizing shapes in their input, but they do not handle specific features such as the temporal evolution of their data. This feature is especially interesting for solar flare prediction since flares are mainly caused by the evolution of active regions illustrated by emerging polarities, energy buildup, and other events. Furthermore, studies using temporal evolution and presenting promising results such as \citet{Guastavino2022} are slowly emerging. In future studies, we will include this feature in our predictions by using LSTM models for solar flare prediction.
\section{Conclusions}
As humanity improves its knowledge and technology, it is also becoming increasingly dependent on it. If our electric infrastructures were to cease to function, it would have a huge impact on society. One of the possible causes of such an event is a strong solar flare. In the past years, there has been increased interest in predicting solar flares in order to prevent such damage. Many studies have been published, and solar flare prediction has become increasingly more accurate, especially with the emergence of DL. Convolutional neural networks can extract features from their input and classify them accurately when correctly used. Moreover, CNNs have been used in solar flare prediction, and every year, increasing progress is made toward a perfect prediction. \\

This study aimed to assess the efficiency of an SPP layer in a CNN for solar flare prediction. This goal was driven by the benefit provided by the SPP layer of keeping the raw data while feeding it to the network instead of losing the information when resizing it. Bypassing the resizing process is especially appealing, as we believe the triggering mechanism of solar flares lies in the small details erased during the usual resizing and preprocessing methods. Consequently, we compared a traditional CNN (Trad-CNN) composed of convolution, max pooling, batch normalization, and fully connected layers with another CNN (SPP-CNN) implementing the SPP layer instead of the max pooling layer before the flattening layer. While the Trad-CNN used a resizing function to fit its input in the network, the SPP-CNN was fed data without any prior manipulation during its evaluation. Furthermore, we also trained our models using a SOL function in the hopes of alleviating the problem of class imbalance and compared it with a BCE loss function.\\

The dataset used was made of SHARP files, cutouts of active regions from magnetograms of the Sun, and it corresponds to the observation of the Sun from May 2010 to August 2021. The goal of our models was to predict flares happening in the next 24 hours using a binary classification. We first trained the models to recognize images of active regions producing a flare in less than 24 hours with a GOES class $\ge$C1.0 and then $\ge$M1.0. Furthermore, through the use of cross-validation, we were able to obtain stable results about the efficiency of the SPP layer. During the evaluation of our models on their test datasets, we observed an overall improvement in the solar flare prediction ability for the SPP-CNN models while classifying images of incoming flares of GOES class $\ge$C1.0 as positive, compared to the traditional CNN. Among the improvements, we note an average increase of 0.1 in TSS, 0.17 in precision, and 0.11 in PR AUC for SPP-CNN models trained using a BCE loss function compared to Trad-CNN models trained with the SOL function. However, there is a drop in precision, TSS, and PR AUC when classifying only images of incoming flares $\ge$M1.0 as positive for all of our models. Introducing C-labeled images into the negative class raised several problems, such as an increased class imbalance and problems linked to the differences and similarities of the magnetic structures of active regions producing intense and mild flares. We also speculate that the SPP-CNN is able to process features inherent in solar flares no matter their GOES class and to be solely focused on these feature during a prediction. Due to the lack of “off-the-shelf” XAI tools, we can only suppose this is due to the process of extracting the maximum value for each tile in the SPP layer and therefore losing a large quantity of information, such as the amount of presence for each feature in tiles. \\

While line of sight magnetograms are simple and easy to use due to their resemblance to images, they provide a limited amount of information about the Sun. In our future studies, we plan to work with different types of observations that may provide more information on solar flare triggering mechanisms, including the SHARP vector magnetograms. Furthermore, one of the main features not handled by CNNs is the temporal evolution of its data. This information is particularly useful for solar flare prediction, and we plan to study its use through LSTM methods.

\begin{acknowledgements}
      We thank the Belgian Federal Science Policy Office (BELSPO) for the provision of financial support in the framework of the Brain-be program under contract number B2/202/P1/DELPHI and in the framework of the PRODEX Programme of the European Space Agency (ESA) under contract number 4000136424. The research that led to these results was also subsidized by Belgian Defence – Royal Higher Institute for Defence (RHID) through contract no 22DEFRA006. This research has made use of NASA’s Astrophysics Data System.
\end{acknowledgements}

\bibliography{paper} 

\begin{appendix}
\onecolumn
\section{Graph of the evolution of the prediction confidence for each model, during the lifetime of an active region}

\begin{figure*}[h!]
   \centering
\includegraphics[width=\linewidth]{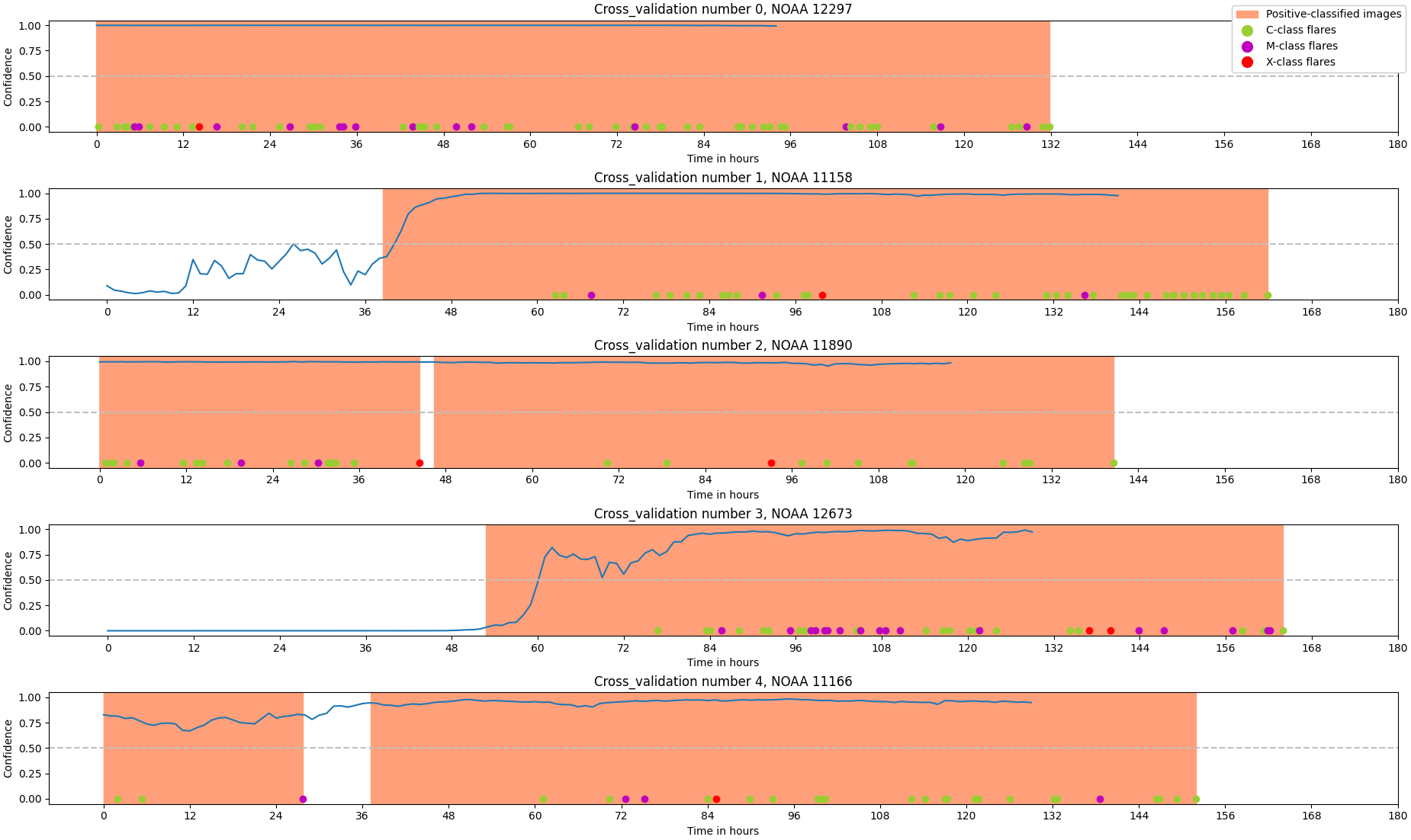}
\caption{Prediction confidence for the prediction of images of active regions producing X class flares by SPP-CNN, trained with BCE loss using the CMX classification. The colored area corresponds to the interval of time during which images are labeled positive in our test dataset.}
          \label{pred_conf_SPP_BCE_CMX_X}%
\end{figure*}
\begin{figure*}[h!]
   \centering
\includegraphics[width=\linewidth]{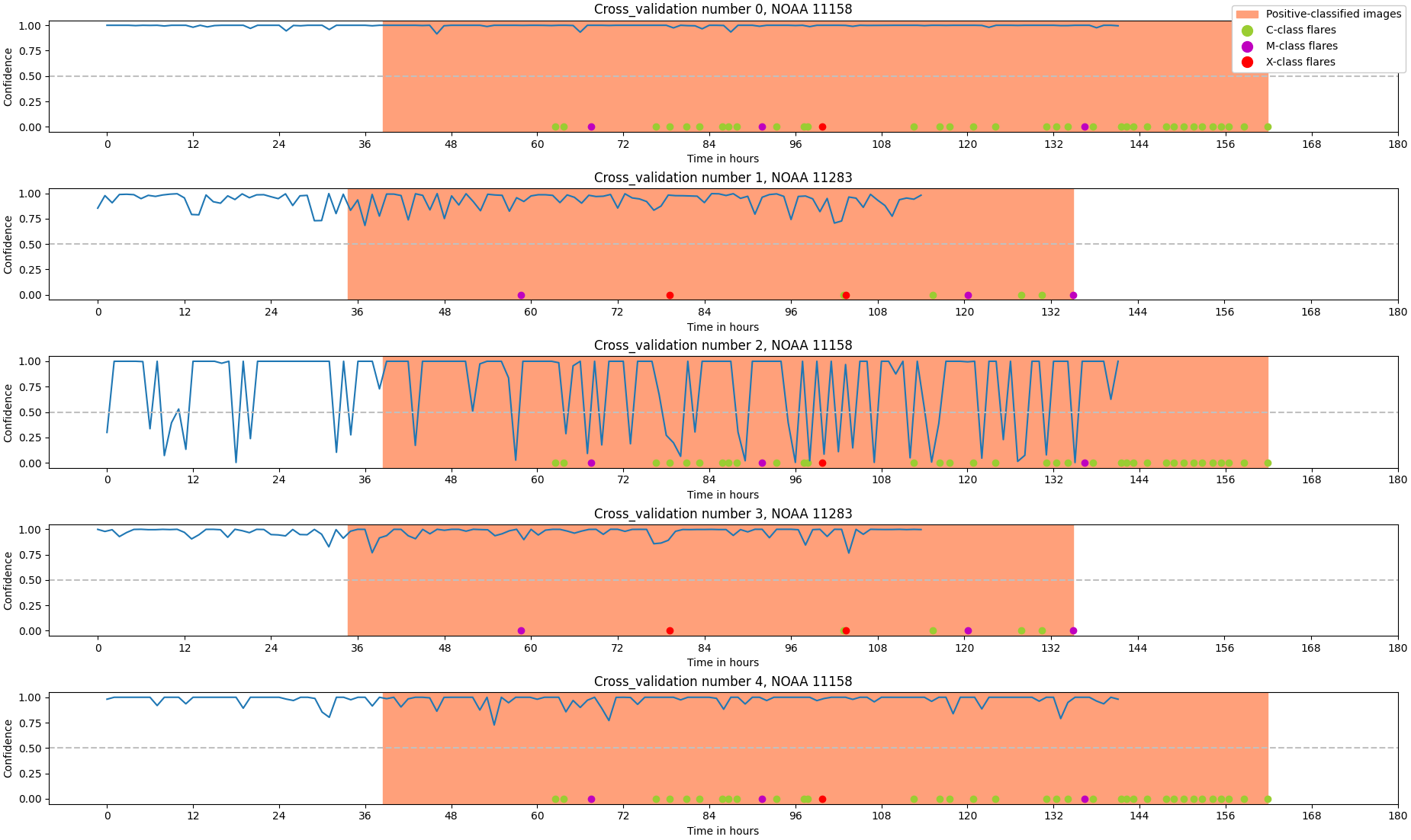}
\caption{Same as Figure \ref{pred_conf_SPP_BCE_CMX_X} but with models using the SPP-CNN architecture and the SOL function.}
          \label{pred_conf_SPP_TSS_CMX_X}%
\end{figure*}
\begin{figure*}[h!]
   \centering
\includegraphics[width=\linewidth]{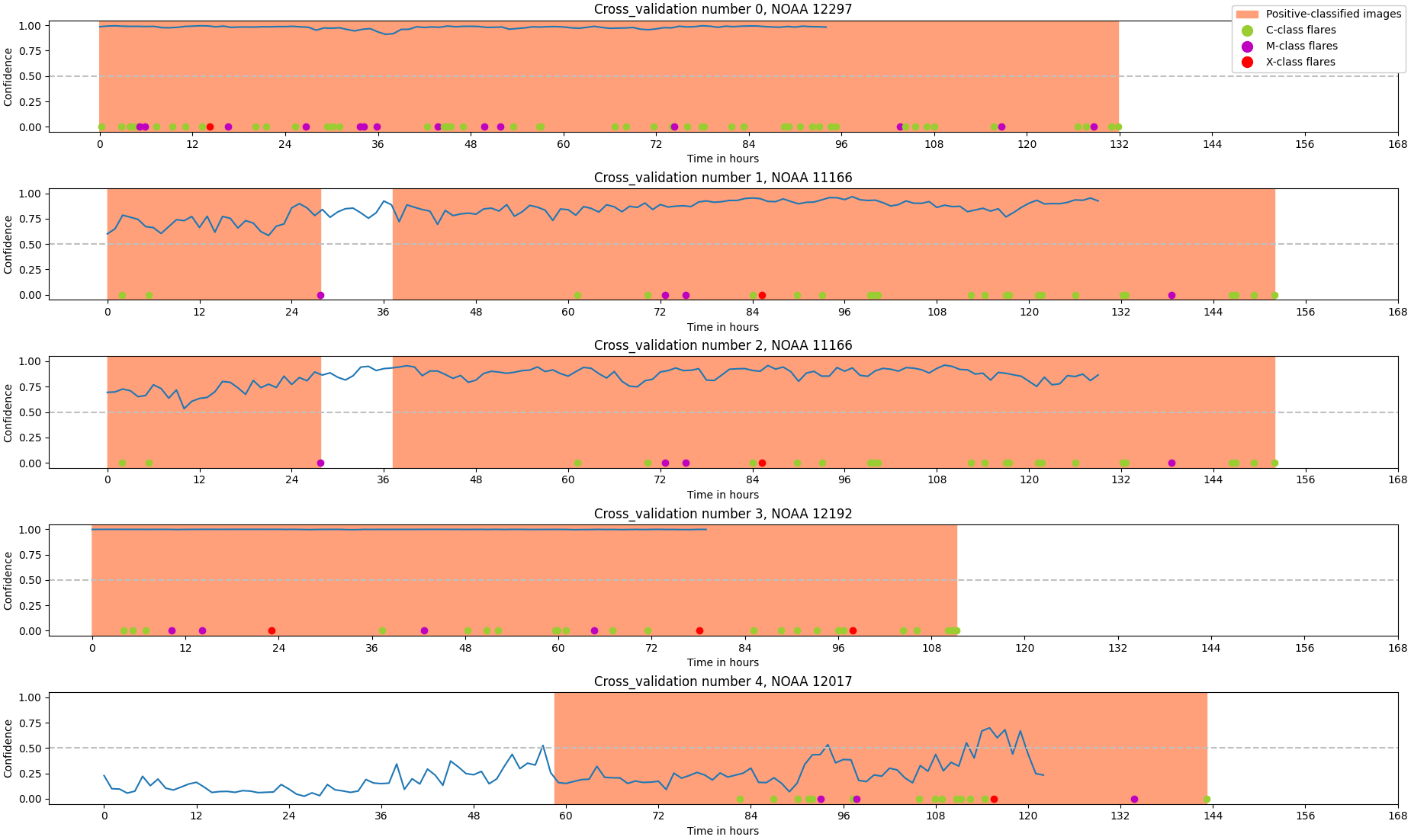}
\caption{Same as Figure \ref{pred_conf_SPP_BCE_CMX_X} but with models using the TRAD-CNN architecture and the BCE loss function.}
          \label{pred_conf_TRAD_BCE_CMX_X}%
\end{figure*}
\begin{figure*}[h!]
   \centering
\includegraphics[width=\linewidth]{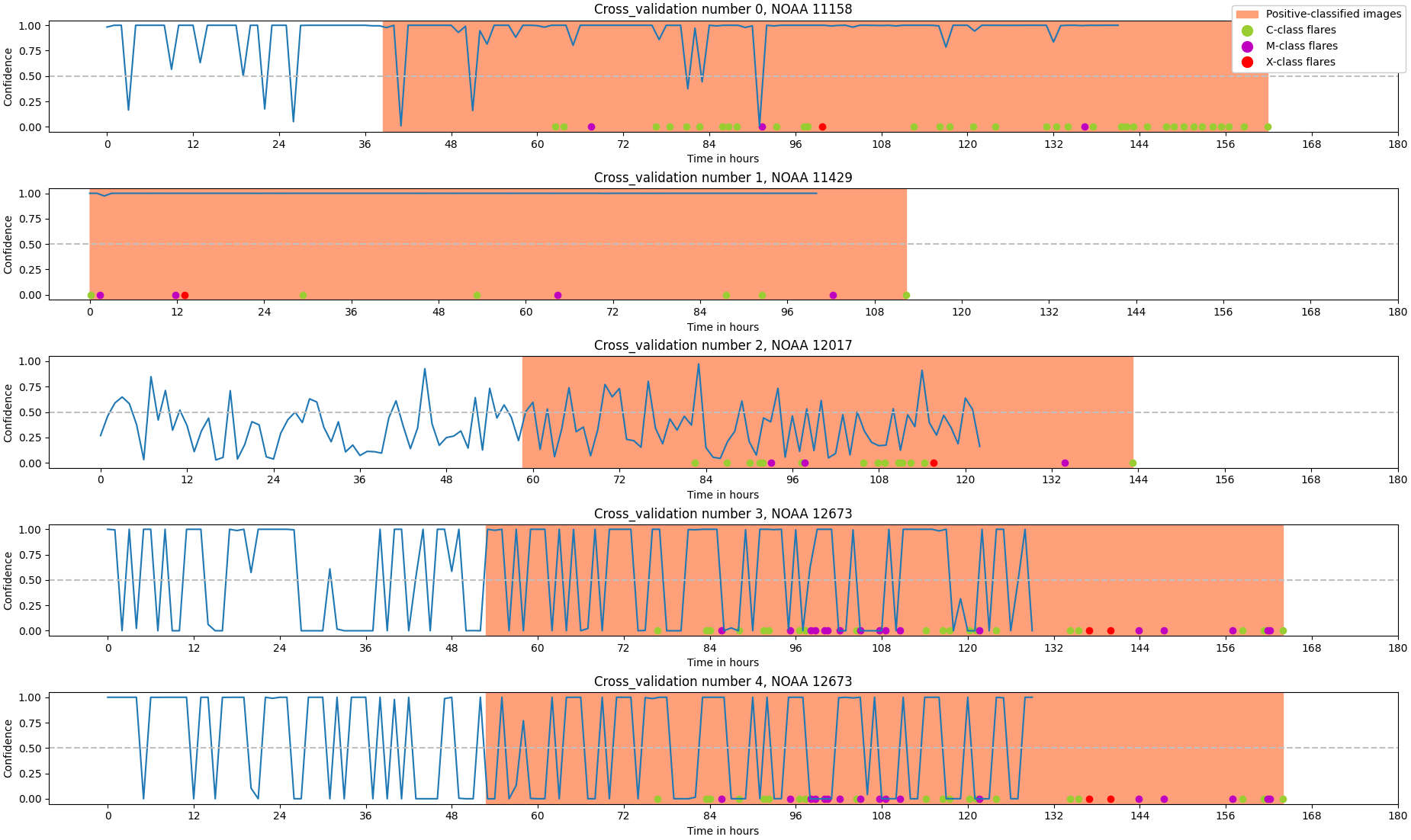}
\caption{Same as Figure \ref{pred_conf_SPP_BCE_CMX_X} but with models using the TRAD-CNN architecture and the SOL function.}
          \label{pred_conf_TRAD_TSS_CMX_X}%
\end{figure*}
\begin{figure*}[h!]
   \centering
\includegraphics[width=\linewidth]{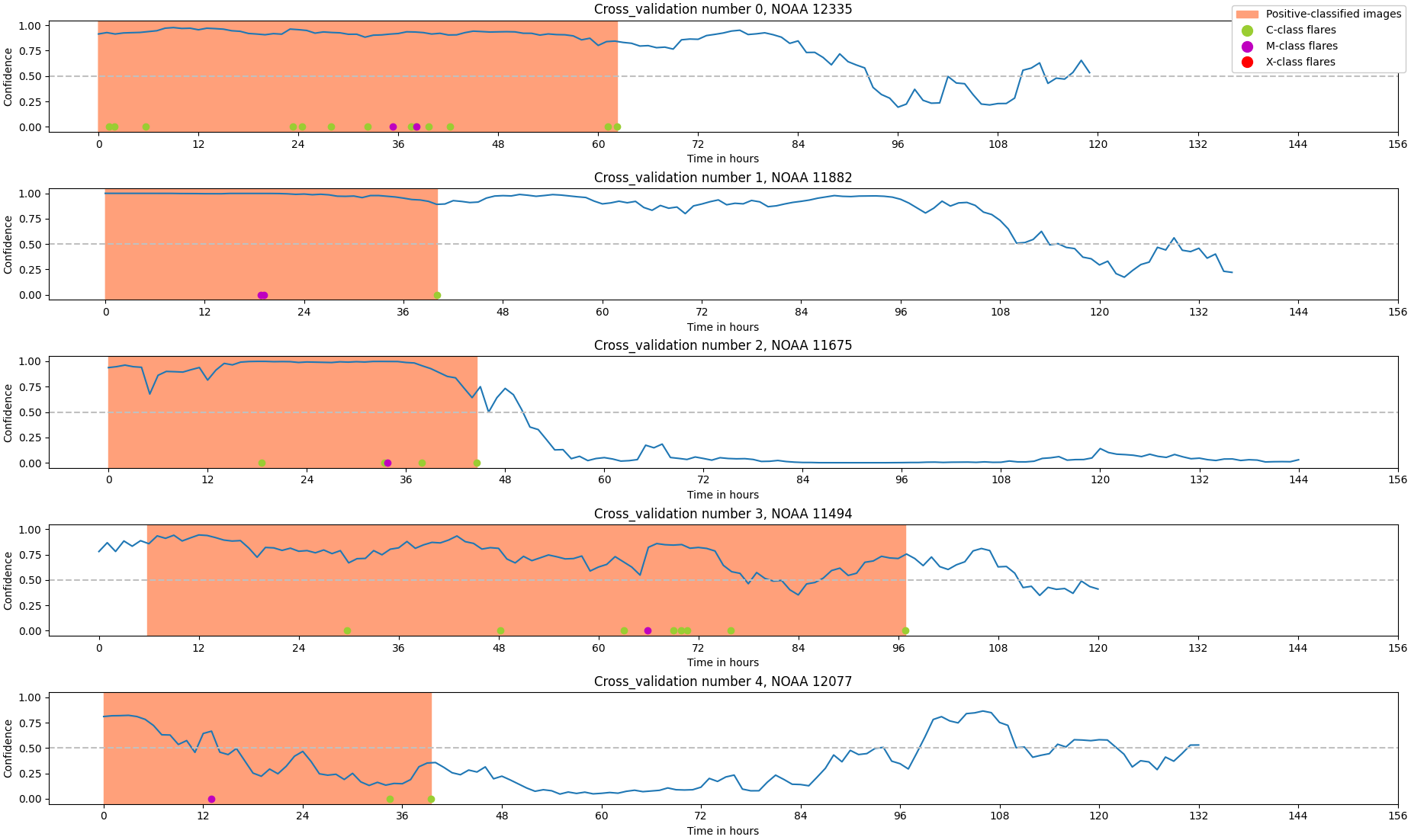}
\caption{Prediction confidence for the prediction of images of active regions producing M class flares as the strongest flare by SPP-CNN, trained with BCE loss using the CMX classification.  The colored area corresponds to the interval of time during which images are labeled positive in our test dataset.}
          \label{pred_conf_SPP_BCE_CMX_M}%
\end{figure*}
\begin{figure*}[h!]
   \centering
\includegraphics[width=\linewidth]{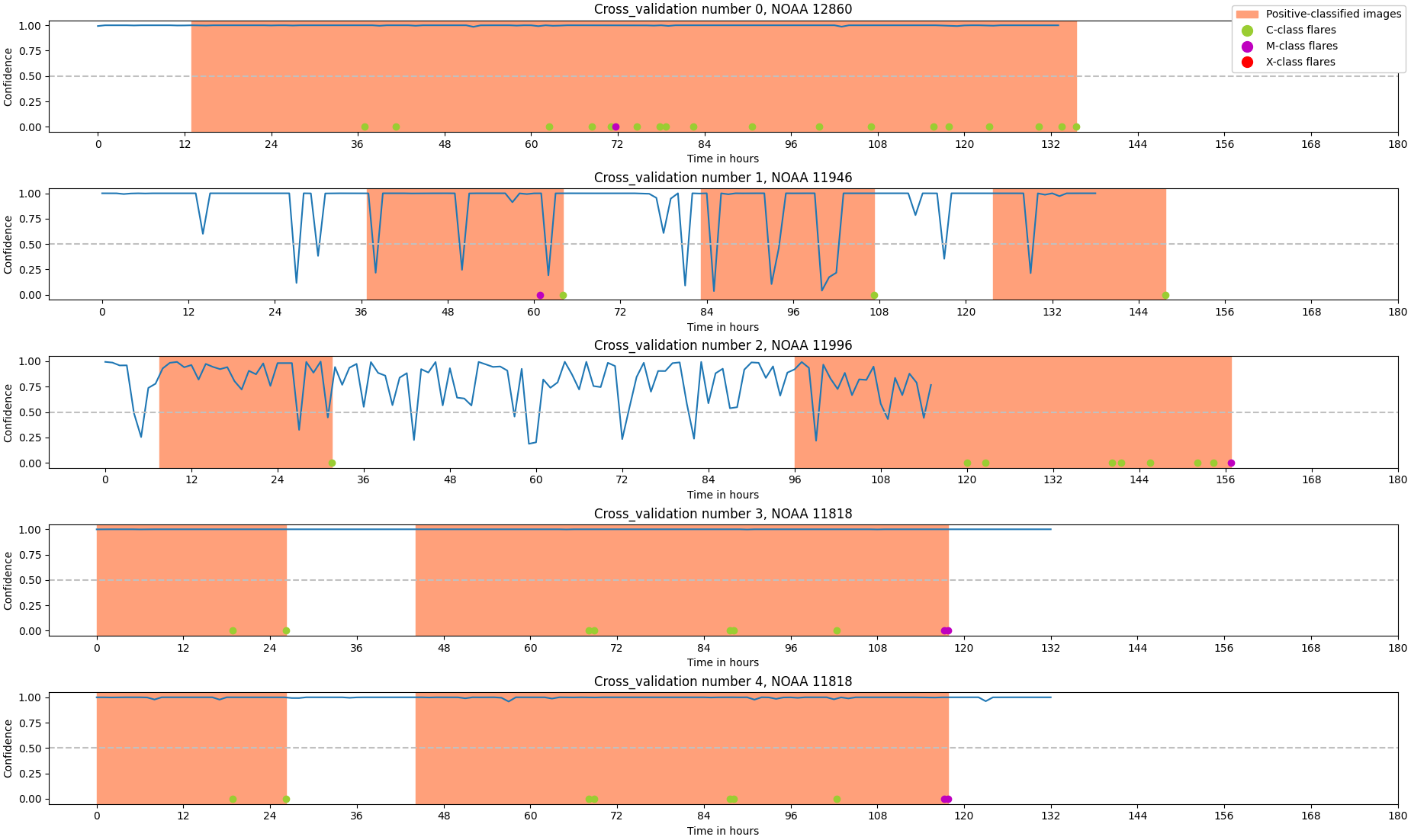}
\caption{Same as Figure \ref{pred_conf_SPP_BCE_CMX_M} but with models using the SPP-CNN architecture and the SOL function.}
          \label{pred_conf_SPP_TSS_CMX_M}%
\end{figure*}
\begin{figure*}[h!]
   \centering
\includegraphics[width=\linewidth]{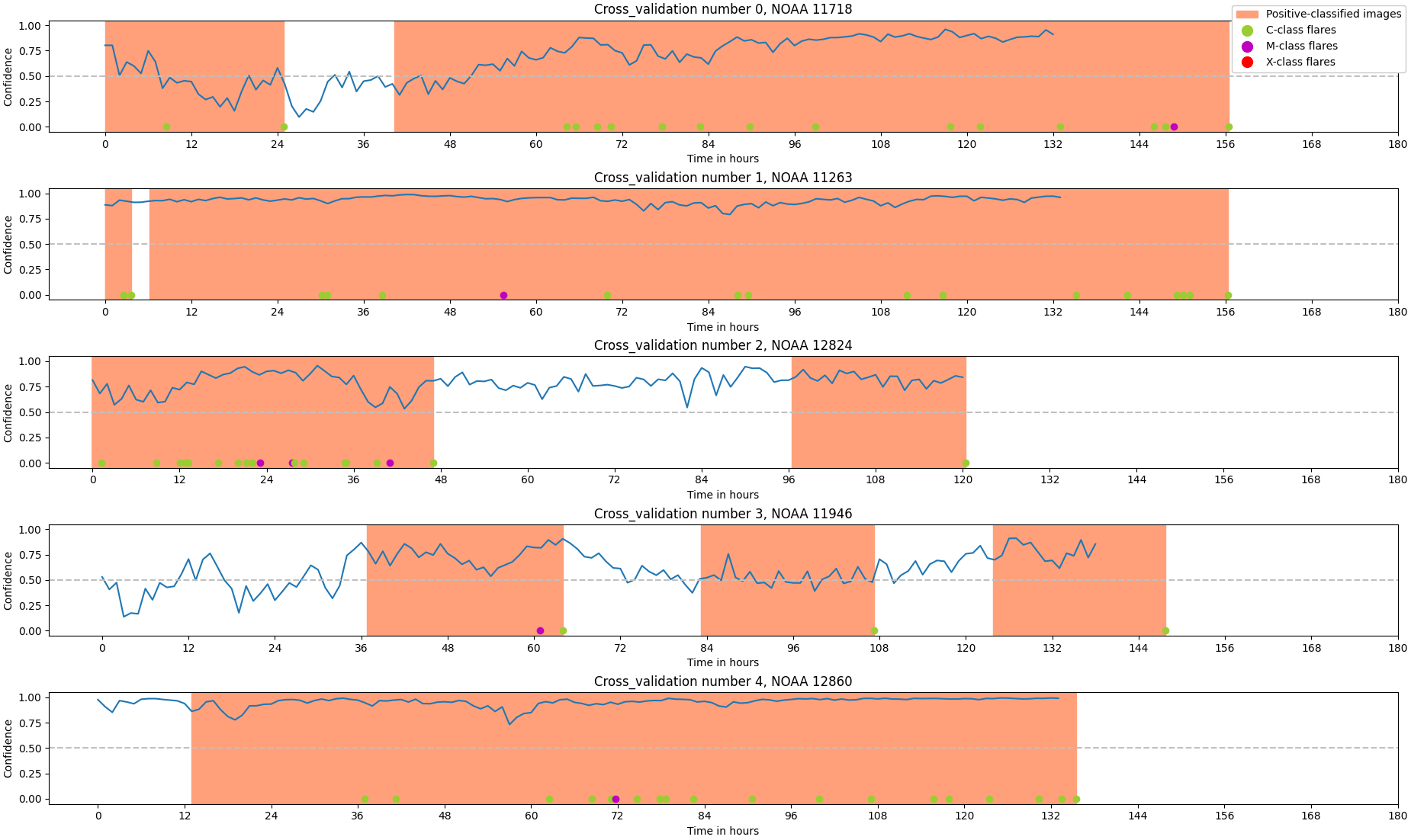}
\caption{Same as Figure \ref{pred_conf_SPP_BCE_CMX_M} but with models using the TRAD-CNN architecture and the BCE loss function.}
          \label{pred_conf_TRAD_BCE_CMX_M}%
\end{figure*}
\begin{figure*}[h!]
   \centering
\includegraphics[width=\linewidth]{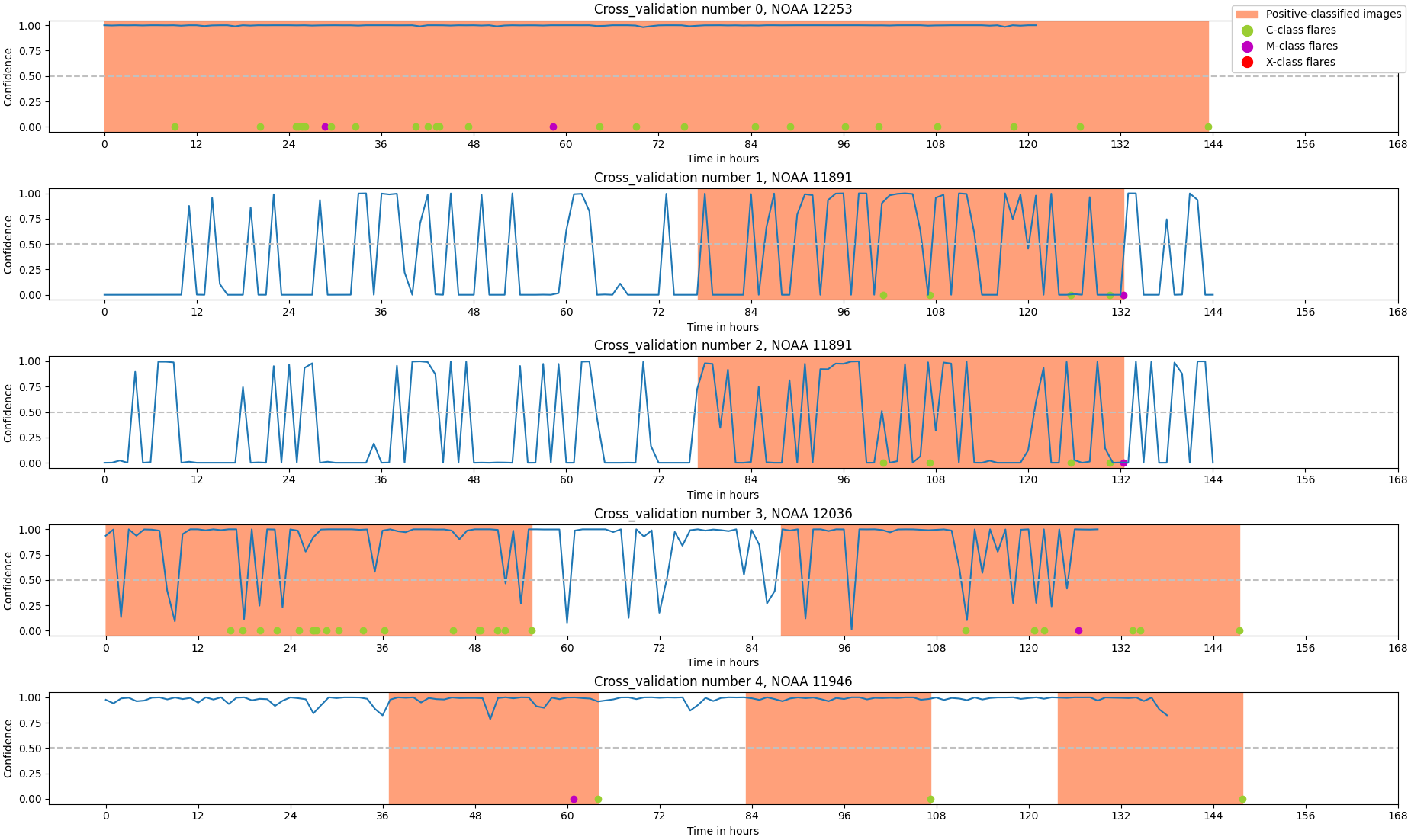}
\caption{Same as Figure \ref{pred_conf_SPP_BCE_CMX_M} but with models using the TRAD-CNN architecture and the SOL function.}
          \label{pred_conf_TRAD_TSS_CMX_M}%
\end{figure*}

\twocolumn
\clearpage

\onecolumn
\section{Tables of label-wise average results of SPP-CNN and Trad-CNN evaluations}

    \begin{table*}[h!]
        \caption{Label-wise average results of SPP-CNN and Trad-CNN evaluations using the CMX classification.}
        \label{table:6}
        \centering
        \begin{tabular}{c c c c c c c c c} 
            \hline\hline 
                Classification & Model & Loss & Class & Label & Correctly Predicted & Wrongly Predicted & Acc\tablefootmark{a}
                & Acc Std\tablefootmark{b} \\
            \hline
            
                CMX & Trad-CNN & BCE & 1 & X & 23 & 4 & 0.83 & 0.33 \\
                    &    &          &    & M & 222 & 18 & 0.92 & 0.03 \\
                    &    &          &    & C & 1064 & 523 & 0.67 & 0.04 \\ \\
                    &    &          & 0  & NF & 7840 & 1020 & 0.89 & 0.02 \\
                    &    &          &    & FX & 19 & 14 & 0.36 & 0.41 \\
                    &    &          &    & FM & 128 & 140 & 0.48 & 0.2 \\
                    &    &          &    & FC & 1596 & 975 & 0.61 & 0.06 \\ \\
 
                    &    & TSS      & 1  & X & 18 & 4 & 0.82 & 0.37 \\
                    &    &          &    & M & 238 & 19 & 0.92 & 0.04 \\
                    &    &          &    & C & 1113 & 478 & 0.69 & 0.08 \\ \\
                    &    &          & 0  & NF & 7883 & 1067 & 0.88 & 0.02 \\
                    &    &          &    & FX & 41 & 34 & \underline{0.68} & 0.29 \\
                    &    &          &    & FM & 1621 & 893 & 0.64 & 0.06 \\
                    &    &          &    & FC & 167 & 124 & 0.59 & 0.1 \\ \\
                
                    &  SPP-CNN  & BCE & 1  & X & 26 & 0 & \underline{1.0} & 0.0 \\
                    &    &          &    & M & 291 & 10 & \underline{0.97} & 0.04 \\
                    &    &          &    & C & 1084 & 453 & \underline{0.71} & 0.15 \\ \\
                    &    &          & 0  & NF & 8449 & 589 & \underline{0.93} & 0.06 \\
                    &    &          &    & FX & 17 & 2 & 0.5 & 0.5 \\
                    &    &          &    & FM & 185 & 75 & \underline{0.76} & 0.15 \\
                    &    &          &    & FC & 1777 & 586 & \underline{0.76} & 0.1 \\ \\
                
                    &    & TSS      & 1  & X & 33 & 0 & 1.0 & 0.0 \\
                    &    &          &    & M & 228 & 23 & 0.9 & 0.06 \\
                    &    &          &    & C & 1112 & 473 & 0.7 & 0.12 \\ \\
                    &    &          & 0  & NF & 6236 & 2446 & 0.72 & 0.12 \\
                    &    &          &    & FX & 7 & 30 & 0.18 & 0.37 \\
                    &    &          &    & FM & 107 & 190 & 0.37 & 0.11 \\
                    &    &          &    & FC & 1483 & 1023 & 0.59 & 0.17 \\ \\
            
        \hline
        \end{tabular}
        \tablefoottext{a}{Accuracy}
        \tablefoottext{b}{Accuracy Standard deviation}
        \tablefoot{Value underlined highlight the best result among all models and Loss function.}
    \end{table*}

    \begin{table*}[h!]
        \caption{Same as \ref{table:6} but using the MX classification.}
        \label{table:7}
        \centering
        \begin{tabular}{c c c c c c c c c} 
            \hline\hline 
                Classification & Model & Loss & Class & Label & Correctly Predicted & Wrongly Predicted & Acc\tablefootmark{a} & Acc Std\tablefootmark{b} \\
            \hline
            
                MX  & Trad-CNN  & BCE & 1  & X & 19 & 7 & 0.73 & 0.38 \\
                    &    &          &    & M & 158 & 86 & 0.66 & 0.13 \\ \\
                    &    &          & 0  & C & 806 & 791 & 0.5 & 0.06 \\
                    &    &          &    & NF & 7811 & 885 & 0.9 & 0.09 \\
                    &    &          &    & FX & 12 & 11 & \underline{0.52} & 0.48 \\
                    &    &          &    & FM & 219 & 70 & \underline{0.75} & 0.16 \\
                    &    &          &    & FC & 2035 & 564 & 0.78 & 0.06 \\ \\

                    &    & TSS      & 1  & X & 15 & 11 & 0.71 & 0.35 \\
                    &    &          &    & M & 227 & 60 & \underline{0.79} & 0.12 \\ \\
                    &    &          & 0  & C & 564 & 920 & 0.37 & 0.22 \\
                    &    &          &    & NF & 5563 & 3445 & 0.63 & 0.27 \\
                    &    &          &    & FX & 28 & 41 & 0.44 & 0.19 \\
                    &    &          &    & FM & 135 & 89 & 0.51 & 0.28 \\
                    &    &          &    & FC & 1330 & 1281 & 0.51 & 0.24 \\ \\
                    
                    & SPP-CNN & BCE & 1  & X & 21 & 5 & \underline{0.78} & 0.39 \\
                    &    &          &    & M & 195 & 75 & 0.69 & 0.39 \\ \\
                    &    &          & 0  & C & 853 & 695 & 0.53 & 0.3 \\
                    &    &          &    & NF & 6780 & 1948 & 0.79 & 0.23 \\
                    &    &          &    & FX & 28 & 5 & 0.24 & 0.42 \\
                    &    &          &    & FM & 189 & 80 & 0.71 & 0.27 \\
                    &    &          &    & FC & 1608 & 883 & 0.65 & 0.3 \\ \\
                
                    &    & TSS      & 1  & X & 19 & 7 & 0.6 & 0.49 \\
                    &    &          &    & M & 154 & 99 & 0.61 & 0.32 \\ \\
                    &    &          & 0  & C & 950 & 542 & \underline{0.64} & 0.21 \\
                    &    &          &    & NF & 7671 & 772 & \underline{{0.91}} & 0.07 \\
                    &    &          &    & FX & 15 & 6 & 0.41 & 0.43 \\
                    &    &          &    & FM & 271 & 137 & 0.68 & 0.28 \\
                    &    &          &    & FC & 1918 & 535 & \underline{0.79} & 0.16 \\ \\
        \hline
        \end{tabular}
        \tablefoottext{a}{Accuracy}
        \tablefoottext{b}{Accuracy Standard deviation}
        \tablefoot{Value underlined highlight the best result among all models and Loss function.}
    \end{table*}

\end{appendix}
\end{document}